\documentclass{article}

\usepackage{authblk}   
\usepackage{hyperref}  
\usepackage{amsmath, amssymb}  
\usepackage{graphicx}  
\usepackage{geometry}  
\usepackage{url,mathtools,graphicx,subcaption}       

\newcommand*\diff{\mathop{}\!\mathrm{d}}
\newcommand\norm[1]{\left\lVert#1\right\rVert}
\DeclarePairedDelimiter\floor{\lfloor}{\rfloor}

\title{Accurate Solutions to Optimal Control Problems via a Flexible Mesh and Integrated Residual Transcription}

\author[1,2]{Lucian Nita\thanks{Corresponding author. Email: \href{mailto:ln416@ic.ac.uk}{ln416@ic.ac.uk}, \href{mailto:lnita@uel.ac.uk}{lnita@uel.ac.uk}}}
\author[1,3]{Eric C.\ Kerrigan}

\affil[1]{Department of Electrical and Electronic Engineering, Imperial College London, London, SW7 2AZ, UK}
\affil[2]{School of Architecture, Computing and Engineering, University of East London, London, E16 2RD, UK}
\affil[3]{Department of Aeronautics, Imperial College London, London, SW7 2AZ, UK}

\date{}

\begin{document}

\maketitle

\renewcommand{\thefootnote}{}
\footnotetext{\textbf{Abbreviations:} OCP, optimal control problem; NLP, nonlinear programming.}
\renewcommand{\thefootnote}{\arabic{footnote}}
\setcounter{footnote}{0}

\begin{abstract}
We propose joining a flexible mesh design with an integrated residual transcription in order to improve the accuracy of numerical solutions to optimal control problems. This approach is particularly useful when state or input trajectories are non-smooth, but it may also be beneficial when dynamics constraints are stiff. Additionally, we implement an initial phase that will ensure a feasible solution is found and can be implemented immediately in real-time controllers. Subsequent iterations with warm-starting will improve the solution until optimality is achieved. Optimizing over the mesh node locations allows for discontinuities to be captured exactly, while integrated residuals account for the approximation error in-between the nodal points. First, we numerically show the improved convergence order for the flexible mesh. We then present the feasibility-driven approach to solve control problems and show how flexible meshing and integrated residual methods can be used in practice. The presented numerical examples demonstrate for the first time the numerical implementation of a flexible mesh for an integrated residual transcription. The results show that our proposed method can be more than two times more accurate than conventional fixed mesh collocation for the same computational time and more than three times more accurate for the same problem size.
\end{abstract}


\section{Introduction}\label{sec:introduction}

Optimal control aims to find ideal input and state trajectories for a dynamical system that will minimize a certain control objective. Dynamics of the physical system are represented by using a set of differential equations typically enforced as path equality constraints. Repeated solving of such problems represents the backbone of modern control techniques such as model predictive control~\cite{rawlings2017model}. There are numerous applications in various fields such as aircraft trajectory planning, control of energy storage systems, financial decision making, and industrial process operations.

Solving optimal control problems is inherently challenging, since an infinite number of decision variables is required to compute the exact solution. Various transcription methods with appropriate mesh refinement techniques have been developed over the years in order to construct increasingly better numerical approximations for the optimal control solutions. However, most methods rely on continuous interpolation functions and struggle to represent potential regions of non-smoothness in a finite-dimensional space. Discontinuity detection is a topic of ongoing research and some alternatives have been proposed in~\cite{liu2015adaptive} and~\cite{paiva2015adaptive}, but they still rely on an a posteriori solution analysis. In this paper, we propose including mesh nodes as part of the decision vector, thus allowing discontinuities to be captured during the solution process, without prior knowledge or assumptions about the solution structure. 

One fundamental difficulty when attempting to numerically solve optimal control problems (OCPs) on a limited-memory, finite-precision processor is the implementation of path constraints. The literature identifies three main solution approaches: dynamic programming, indirect methods, and direct methods. In this paper, we will focus our attention on direct transcription methods that aim to discretize the OCP first and then to find the optimal solution of the discrete problem. The most popular transcription methods are direct collocation and multiple shooting~\cite{diehl2006fast}. Direct collocation~\cite{kelly2017introduction} relies on function approximation using support points to generate a large and sparse optimization problem. The method is typically well suited for complex problems with path constraints, but the solution accuracy heavily depends on the number of the discretization points. Hence, direct collocation would struggle to accurately capture the dynamics constraints using coarse meshes. In our approach, we try to mitigate this drawback while maintaining the benefits of direct transcription methods. Thus, our work employs a state-of-the-art integrated residual transcription method that generalizes direct collocation by decoupling the support points from the evaluation points.

The idea of looking for an optimal mesh has been explored since the 1970s in the works of~\cite{babuvvska1978error}, and several recent works proposed various discontinuity capturing techniques, as well as adaptive mesh refinement procedures~\cite{miller2021mesh,liu2015adaptive}. The drawback of existing work is that most rely on direct collocation transcription. However, none of the methods explore the option of using a flexible mesh alongside more modern transcription methods based on integrated residuals. As will be shown in this paper, finding an optimal mesh for a collocation scheme can be problematic, since there is no control over the approximation error between the mesh nodes. Furthermore, the majority of current methods view mesh design as a post-processing task handled during mesh refinement, and consequently find it challenging to deliver a straightforward solution that can be applied immediately. Our aim is to first provide a feasible solution that accurately enforces dynamics constraints over the entire time domain.  

In this paper, our aim is to propose a novel strategy for mesh design used as part of a cutting-edge transcription method for optimal control and compare it against a commonly used method, namely direct collocation. We extend the work in~\cite{9684961} by introducing a flexible mesh with nodes placed at optimal locations, and complement the work in~\cite{nita2022solving} and~\cite{nita2022fast} by comparing our solution approach to traditional collocation. The main contributions of this paper are: 
\begin{enumerate}
    \item We formulate the discretized version of the original control problems with mesh nodes as decision variables (flexible mesh) to achieve an optimal mesh design.
    \item We explore the relevance of the chosen transcription method and show why a flexible mesh would work better with an integrated residual method compared to direct collocation. 
    \item We propose a two-stage feasibility driven approach returning a feasible point in case of early termination. The idea is well suited for real time model predictive control and safety critical control applications.
    \item We show why a flexible mesh is beneficial for stiff and discontinuous systems.
    \item We demonstrate the effectiveness of this strategy using two numerical examples: a Van-der-Pol singular control formulation with a discontinuous optimal solution and a Two-Link Robot Arm application with a fully continuous optimal input.
\end{enumerate}   

The remainder of the paper is organized as follows: In Section~\ref{sec:ch2}, we describe the problem formulation and introduce the notation used. In Section~\ref{sec:ch3} we present our novel solution method and showcase its benefits on illustrative examples. Section~\ref{sec:3A} talks about the optimal mesh design, while in Section~\ref{sec:3B} we introduce the integrated residual method as an extension of the collocation method, and explain why using flexible meshing with collocation might not yield the best outcomes. The general solution method that ensures feasibility in time-critical applications is introduced in Section~\ref{sec:3C}, and some remarks on mesh refinement and algorithmic implementation are given in Section~\ref{sec:3D}. In Section~\ref{sec:ch4}, two more involved optimal control examples are provided to highlight the effectiveness of the newly proposed method. Section~\ref{sec:ch5} concludes with a summary of the main achievements, underlines the relevance of these findings, and proposes future research directions.

\section{Preliminaries}\label{sec:ch2}
\subsection{Problem formulation}\label{sec:ch2A}

The objective functional of many optimal control and estimation problems can be written in the general Bolza form
\begin{equation}
    \label{eq:eqn1}
    J = \phi(x(t_0),x(t_f),t_0,t_f) + \int_{t_0}^{t_f}{L(x(t),u(t),t)} \diff t,
\end{equation}
where  $x : \mathbb{R} \rightarrow \mathbb{R}^{N_x}$ are the state variables and are assumed to be continuous, $\dot{x} : \mathbb{R} \rightarrow \mathbb{R}^{N_x}$  are the time derivatives of the state $x$, and $u : \mathbb{R} \rightarrow \mathbb{R}^{N_u}$ are the control inputs. $\phi:\mathbb{R}^{N_x} \times \mathbb{R}^{N_x} \times \mathbb{R} \times \mathbb{R} \rightarrow \mathbb{R}$ is the Mayer cost functional, also called the boundary cost, with $t_0\in\mathbb{R}$ and $t_f\in\mathbb{R}$ being the initial and final simulation times, respectively. $L:\mathbb{R}^{N_x} \times \mathbb{R}^{N_u} \times \mathbb{R} \rightarrow \mathbb{R}$ is the Lagrange cost functional, typically called the path cost. 

The problem we are aiming to solve can be formulated as
\begin{subequations}
    \label{eq:equation2}
    \begin{align}
        \min_{x(\cdot),u(\cdot),t_0,t_f} & J(x(\cdot),u(\cdot),t_0,t_f) \label{eq:equation2A}\\
        \textrm{s.t.} \quad & F(\dot{x}(t),x(t),u(t),t)=0 \quad \forall t \in[t_0,t_f], \label{eq:equation2B}\\
        & G(\dot{x}(t), x(t), u(t), t) \leq 0 \quad \forall t \in [t_0,t_f], \label{eq:equation2C}\\
        &\Psi_E(x(t_0), x(t_f), t_0, t_f) = 0, \label{eq:equation2D}\\
        &\Psi_I(x(t_0), x(t_f), t_0, t_f) \leq 0, \label{eq:equation2E}
    \end{align}
\end{subequations}
 where the function $F : \mathbb{R}^{N_x} \times \mathbb{R}^{N_x} \times \mathbb{R}^{N_u} \times \mathbb{R} \rightarrow \mathbb{R}^{N_F}$, describes the dynamical model of the system. $G : \mathbb{R}^{N_x} \times \mathbb{R}^{N_x} \times \mathbb{R}^{N_u} \times \mathbb{R} \rightarrow \mathbb{R}^{N_G}$ defines $N_G$ path inequality constraints. In this paper, we will focus on accurately representing the dynamic constraints~\eqref{eq:equation2B} that define a set of $N_F$ equality constraints that must be satisfied by the controlled system. Most well-defined problems will contain boundary equality constraints that are represented by $\Psi_E : \mathbb{R}^{N_x} \times \mathbb{R}^{N_x} \times \mathbb{R} \times \mathbb{R} \rightarrow \mathbb{R}^{N_E}$, with a common example being $t_0=0$. Furthermore, the boundary inequality constraints are contained in $\Psi_I : \mathbb{R}^{N_x} \times \mathbb{R}^{N_x} \times \mathbb{R}  \times \mathbb{R} \rightarrow \mathbb{R}^{N_I}$. In free end-time problems, one may encounter constraints of the form $t_l \leq t_f \leq t_u$ where $t_l$ and $t_u$ are lower and upper bounds for the final time.

\subsection{Meshing and notation}\label{sec:ch2B}

In order to obtain a numerical solution to~\eqref{eq:equation2}, most interpolation-based methods rely on approximating the state $x(\cdot)$ and input $u(\cdot)$ trajectories using a set of piecewise continuous interpolation functions $\tilde{x}:\mathbb{R}\to\mathbb{R}^{N_x}$ and $\tilde{u}:\mathbb{R}\to\mathbb{R}^{N_u}$, which are called basis functions or approximation functions. $\tilde{x}$ and $\tilde{u}$ are parameterized by a finite number of decision variables $s_i^j$, $c_i^j$ (where subscript $i$ is the subdomain index and superscript $j$ refers to the support point index). The core idea behind most conventional methods is that the numerical solution would approach the exact analytical solution as the number of decision variables increases. 

If the state and input functions $x(\cdot)$ and $u(\cdot)$ are known to be continuous and there is no uncertainty present, one might attempt to represent the numerical solutions $\tilde{x}(\cdot)$ and $\tilde{u}(\cdot)$ as a single polynomial of increasing degree. This approach is termed a $p$-method and generally fails to provide precise approximations in many real-world situations if the solution is not smooth. To mitigate the drawbacks of $p$-methods and be able to accurately represent regions of interest where rapid changes may occur, most algorithms resort to subdividing the time domain $[t_0, t_f]$ into $N$ subdomains (i.e.\ subintervals $[t_i,t_{i+1}]$) such that 
    \begin{subequations}
    \label{eq:equation3}
    \begin{align}
    &\mathcal{T}_i:= [t_{i}, t_{i+1}]  \subset[t_0,t_f], \ \forall i \in \{0, \dots,N-1\},\label{eq:equation3A}\\
    &\cup_{i=0}^{N-1} [t_i,t_{i+1}]=[t_0,t_f],\label{eq:equation3B}\\
    &t_i < t_{i+1},\ \forall i \in \{0, \dots,N-1\},\label{eq:equation3C} 
    \end{align}
    \end{subequations}
where $t_{N}=t_{f}$. To keep the problem tractable and still ensure accurate solutions, it is necessary to use several smaller subintervals for areas with steep gradients, while fewer and larger subintervals are preferred elsewhere. Increasing the number of intervals $N$ is equivalent to decreasing the average length of the subinterval, historically denoted by $h$. Therefore, varying $N$ while maintaining a fixed polynomial degree to approximate the state and input functions within each subdomain is known as the $h$ method. Most modern mesh refinement techniques rely on varying both subdomain size and polynomial degree and are called $hp$-methods.

$\tilde{x}(\cdot)$ and $\tilde{u}(\cdot)$ are usually chosen to be piecewise continuous (continuous inside each interval $[t_i, t_{i+1}]$), so inner interpolation meshes with internal supports $\tau_i^j$ for $\tilde{x}(\cdot)$ and $\mu_i^j$ for $\tilde{u}(\cdot)$ are required, where $\tau_i^j$ denotes the $j^{\text{th}}$ nodal point inside $i^{\text{th}}$ interval. $s_i^j$ and $c_i^j$ are the values of the state and input approximations when evaluated on the internal supports (i.e.\ $\tilde{x}(\tau_i^j)=s_i^j$ and $\tilde{u}(\mu_i^j)=c_i^j$). Direct transcription methods rely on a linear combination of $s_i^j$ and $c_i^j$ to construct polynomial approximations $\chi_i$ and $\xi_i$ of degrees $a$ and $b$, respectively, within each subdomain $[t_i,t_{i+1}]$ such that for all $i\in\{0, \dots,N-1\}$:
    \begin{subequations}
    \label{eq:equation4}
    \begin{align}
        \tilde{x}(t)&=\chi_i(t)
        :=\frac{\sum_{j=0}^{a}{\frac{m_i^j(t)}{ w_i^j}\cdot s_i^j}}{\sum_{j=0}^{a} \frac{m_i^j(t)}{ w_i^j}},\ \forall t\in[t_i,t_{i+1}]\label{eq:equation4A}\\
        \tilde{u}(t)&=\xi_i(t)
        :=\frac{\sum_{j=0}^{b}{\frac{n_i^j(t)}{ v_i^j}\cdot c_i^j}}{\sum_{j=0}^{b} \frac{n_i^j(t)}{v_i^j}},\ \forall t\in[t_i,t_{i+1}] \label{eq:equation4B}
    \end{align}
    \end{subequations}
where $s_i^j=\chi_i(\tau_i^j)=\tilde{x}(\tau_i^j)$ and $c_i^j=\xi_i(\mu_i^j)=\tilde{u}(\mu_i^j)$ are nonlinear program (NLP) decision variables, and $\tau_i^j$ and $\mu_i^j$ are interpolation nodes. The interpolating functions $m_i^j(t)= \prod_{l=0,l\neq j}^{a} (t-\tau_i^l)$ and $n_i^j(t)= \prod_{l=0,l\neq j}^{b} (t-\mu_i^l)$ are used to evaluate the state and input approximations $\tilde{x}(\cdot)$ and $\tilde{u}(\cdot)$ between the support nodes $\tau_i^j$ and $\mu_i^j$, while $w_i^j$ and $v_i^j$ are the corresponding polynomial weights $w_i^j=\prod_{l=0,l\neq j}^{a}(\tau_i^j-\tau_i^l)$, $v_i^j=\prod_{l=0,l\neq j}^{b}(\mu_i^j-\mu_i^l)$. Various choices for $m_i^j(\cdot)$, $n_i^j(\cdot)$ and their corresponding weights, as well as support point distributions $\tau_i^j$ and $\mu_i^j$ are presented in~\cite{berrut2004barycentric}. 

\begin{figure}
    \centering
    \includegraphics[width=0.7\columnwidth]{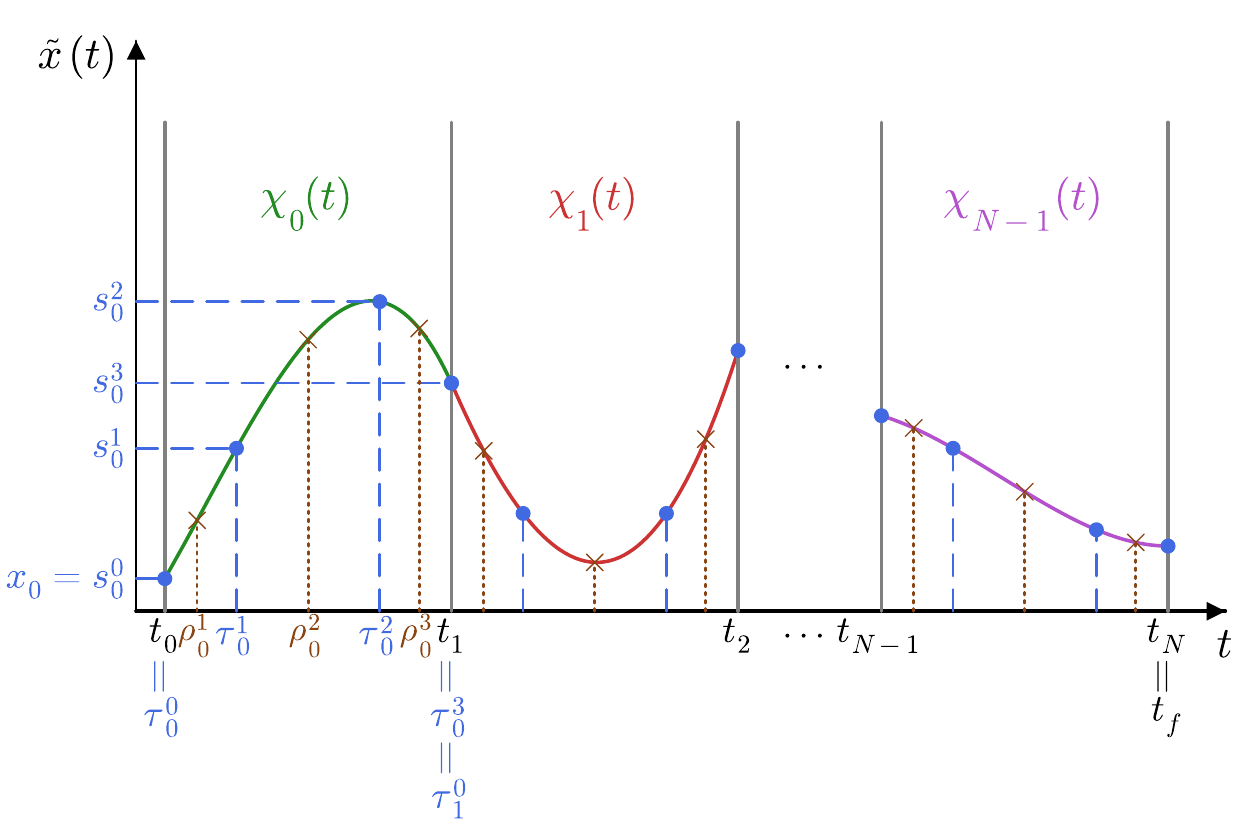}
    \includegraphics[width=0.7\columnwidth]{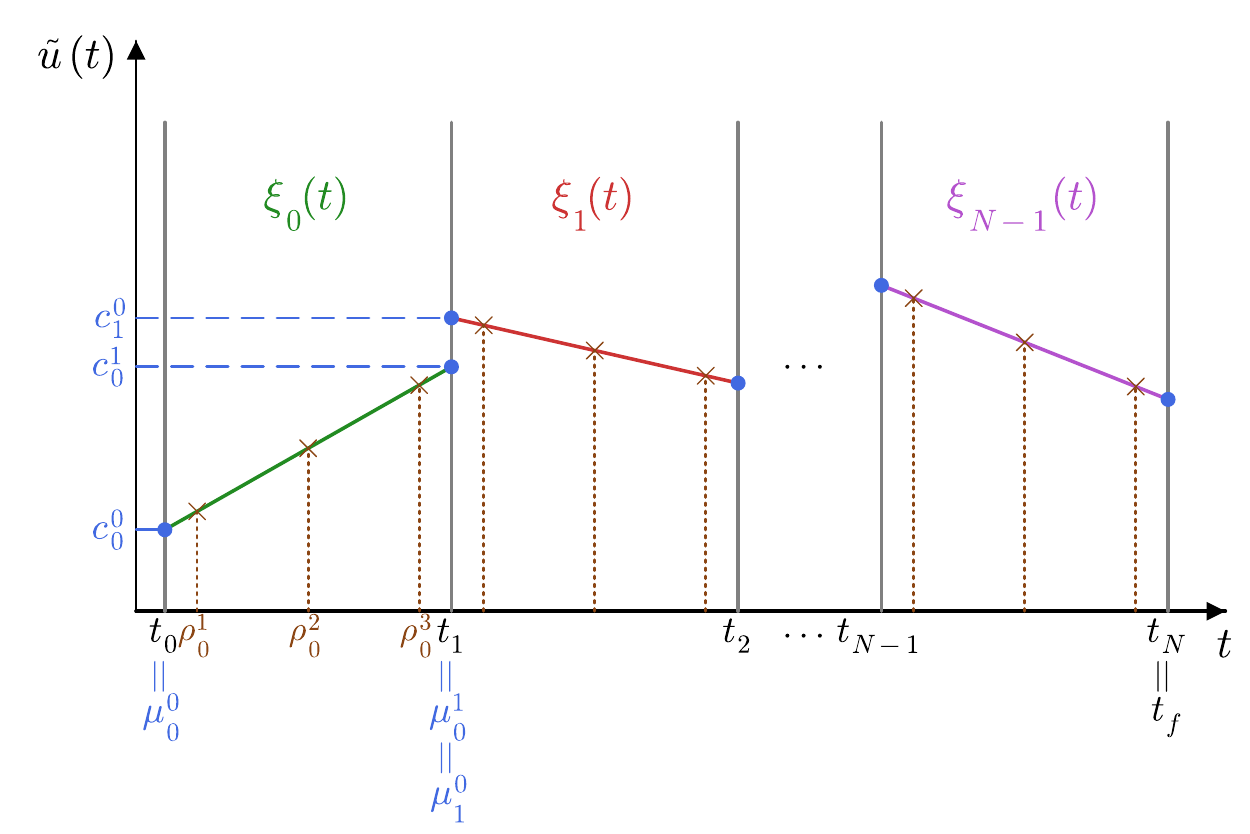}
    \caption{Meshing  for state and input approximations $\tilde{x}$ and $\tilde{u}$ used in constructing numerical solutions for~\eqref{eq:equation2}. Interpolation mesh is denoted by $\tau_i^j$ and $\mu_i^j$, quadrature mesh is $\rho_i^k$ and decision variables are $s_i^j$ and $c_i^j$. }
    \label{fig:figure1}
\end{figure}

To better illustrate these concepts along with the notation used, Figure~\ref{fig:figure1} shows a potential state approximation $\tilde{x}(\cdot)$ constructed using piecewise cubic polynomials ($a=3$) over the support mesh $\tau_i^j$ highlighted in light blue, and a potential input approximation $\tilde{u}(\cdot)$ constructed using piecewise linear functions ($b=1$) over a different support mesh $\mu_i^j$. The state trajectories are assumed to be continuous, with continuity constraints enforced by using the same decision variable $s_{i}^a$ to represent both $\tilde{\chi}_i(t_{i+1})$ and $\tilde{\chi}_{i+1}(t_{i+1})$, while the input trajectories can be discontinuous at the mesh nodes $t_i$. This assumption is necessary in order to be able to capture potentially discontinuous input trajectory curves, but one can enforce input continuity by enforcing constraints $c_i^b=c_{i+1}^0$ for all $i\in\{1,\dots,N-1\}$, or by using the same NLP variable to represent both $c_i^b$ and $c_{i+1}^0$, thus reducing the size of the problem. When required, most conventional transcription methods perform numerical integration using internal supports as quadrature nodes. However, they do not necessarily have to coincide. An evaluation mesh $\rho_i^j$ (plotted in light brown in Figure~\ref{fig:figure1}) will be used to calculate the quadrature as described in Section~\ref{sec:ch4}. In Figure~\ref{fig:figure1} we have used the Chebyshev extreme points to produce the internal node distribution and the Legendre zero points for the evaluation mesh.

\section{Solution method}\label{sec:ch3}

Existing solution methods have a few key limitations:
\begin{itemize}
    \item Inability to provide a satisfactory performance if the functions to be approximated $x(\cdot)$ and $u(\cdot)$ are not smooth.
    \item Failure to account for the inter-nodal error at solve time. 
    \item The returned solution after each mesh refinement iteration may not always be feasible.
\end{itemize}
In this section, we aim to address these limitations and propose a novel solution method for solving optimal control problems.

\subsection{Flexible mesh design}\label{sec:3A}

The transcription process requires the construction of various different meshes and sampling locations, all used for different purposes. Before attempting to solve an optimal control problem numerically, one has to make a number of decisions regarding where best to place various mesh nodes. As explained previously, most modern methods rely on subdividing the domain into multiple subintervals $[t_i,t_{i+1}]$. However, where to best place the nodal points $t_i$ is an open question. Most practical solvers assume the starting mesh is equispaced (i.e.\ $t_{i+1}-t_i=h,\, \forall i\in\{0,\dots,N-1\}$ where~$h$ is a constant defined a priori). More complex alternatives explore various other distributions and the mesh may not remain equispaced after a few refinement iterations (as more mesh nodes will be added inside some subintervals while other subintervals will not be altered), but conventional methods assume that the mesh is known and fixed at solve time.    

In our proposed idea, the mesh nodes $t_i$ will become decision variables and will be included as part of the NLP problem formulation. This is what will be referred to as the \emph{flexible mesh}. 

\subsubsection{Flexible mesh for function approximation}\label{sec:3AA}

\begin{figure}
    \centering
    \includegraphics[width=0.8\columnwidth]{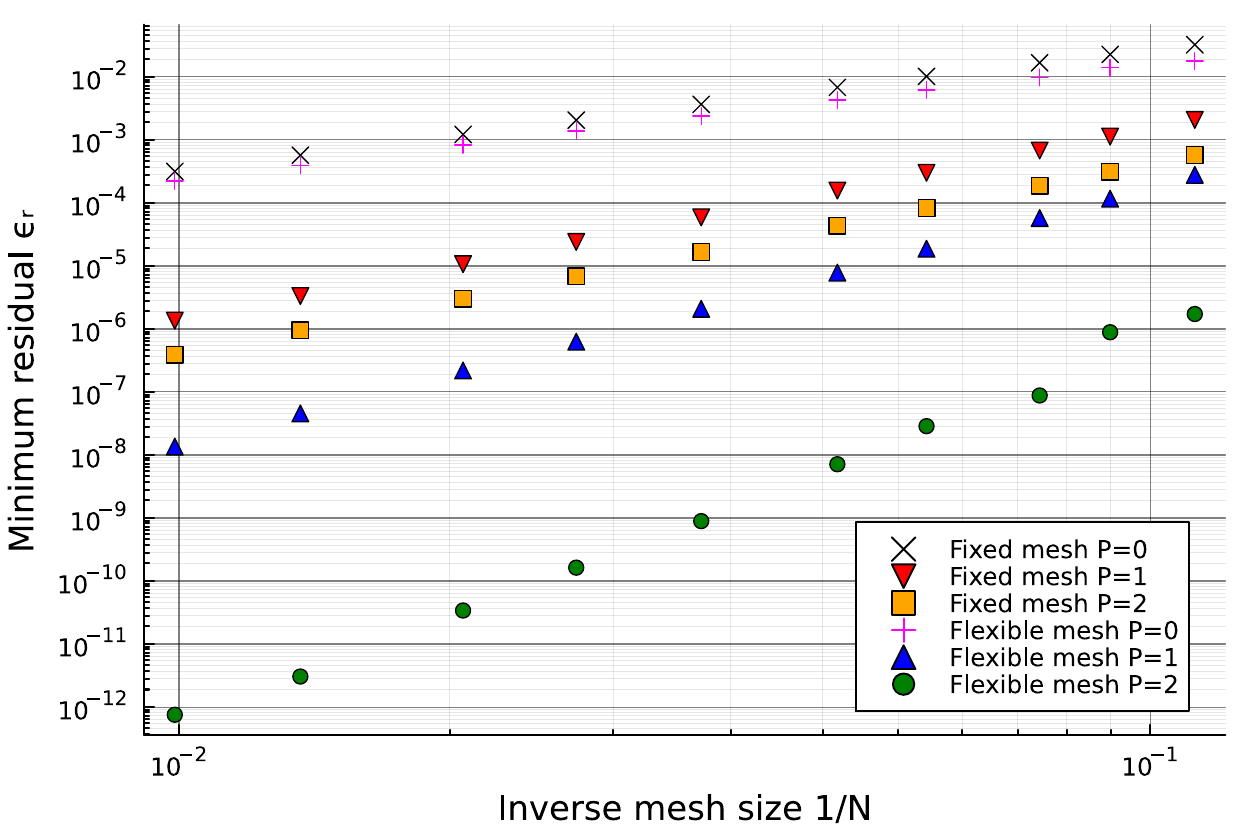}
    \caption{Order of convergence plot for approximating $u(t)=|\cos(\pi t)|$ for fixed vs flexible mesh as polynomial degree $P$ is increased for various mesh size $N$ between $10$ and $100$. The residual function is $\epsilon_r=\int_0^2 (\tilde{u}(t)-u(t))^2 dt$.}
    \label{fig:figure2}
\end{figure}

As a particular case of our numerical optimal control solver, one can use a flexible mesh just for function approximation, making the number of state variables $N_x=0$ and only having path constraints $F(\dot{x}(\cdot),x(\cdot),u(\cdot),\cdot)$ for control inputs $u(\cdot)$. To understand the advantages of a flexible mesh, let us explore the example of least squares fitting of a known curve $u(t)-|\cos(\pi t)|=0, \forall t\in[0,2]$ where $|\cdot|$ represents the absolute value function (i.e.\ finding $\tilde{u}(\cdot)$ such that $\tilde{u}(t)-|\cos(\pi t)|=\tilde{u}(t)-u(t)\approx 0,\, \forall t \in [0,2]$). By construction $u(t)$ is $C^0$ continuous, but non-smooth, with discontinuous derivative at $t=0.5$ and $t=1.5$. The convergence order will be limited by the best polynomial approximation of the same degree as $u(\cdot)$. In other words, if $u(\cdot)$ is $C^n$ continuous, the convergence order of an $hp$-method aiming to approximate $u(\cdot)$ with a degree $P$ piecewise polynomial $\tilde{u}(\cdot)$ will be capped at $P=n+1$ unless some mesh nodes $t_i$ align with the points of non-smoothness. The numerical results presented in Figure~\ref{fig:figure2} show that for a fixed mesh, the convergence order (i.e.\ the slope in a log-log graph) no longer increases above $P=1$. However, one can notice that the total discretization error $\epsilon_r=\int_0^2 (\tilde{u}(t)-u(t))^2 dt$ is smaller when using a flexible mesh. Additionally, the convergence order still increases for $P>1$, which highlights that points of non-smoothness are successfully captured by the flexible mesh.  

\subsubsection{Flexible mesh for feasibility problems and differential equations} 
\label{sec:3AB}

A flexible mesh can be useful for solving non-smooth differential equations ($N_u=0$) and constrained control problems ($J$ is constant and the aim is constraint satisfaction). To find a point satisfying the dynamic constraints, one may attempt to solve 
\begin{subequations}
    \label{eq:feas_cont}  
    \begin{align}
        \min_{x(\cdot),u(\cdot)} \quad & \int_{t_0}^{t_f}\norm{F(\dot{x}(t),x(t),u(t),t)}_2^2 dt, \label{eq:5a}\\
        \textrm{s.t.} \quad & \eqref{eq:equation2C}, \eqref{eq:equation2D}, \eqref{eq:equation2E}. \label{eq:5b}
    \end{align}
\end{subequations}

After discretization using a flexible mesh, the corresponding NLP takes the form of a nonlinear least-squares problem 
\begin{subequations}
    \label{eq:diff_eq_flex_pb_def}  
    \begin{align}
        \min_{\mathbf{s},\mathbf{c},\mathbf{t}} \quad & \sum_{i=0}^{N-1}{\sum_{k=1}^{Q}{\sigma_i^k \cdot \norm{ F(\dot{\tilde{x}}(\rho_i^{k}),\tilde{x}(\rho_i^{k}),\tilde{u}(\rho_i^{k}),\rho_i^k)}^2_2}}, \label{eq:deqA}\\
        \textrm{s.t.} \quad & G(\dot{\tilde{x}}(\tau_i^j), \tilde{x}(\tau_i^j), \tilde{u}(\tau_i^j), \tau_i^j) \leq 0 \quad \forall \tau_i^j \in \tau, \label{eq:deqB}\\
        &\Psi_E(\tilde{x}(t_0), \tilde{x}(t_N), t_0, t_N) = 0, \label{eq:deqC}\\
        &\Psi_I(\tilde{x}(t_0), \tilde{x}(t_N), t_0, t_N) \leq 0, \label{eq:deqD}\\
        &t_{i+1}-t_i\geq \frac{t_{tol}}{N} \quad \forall i \in \{0,\dots,N-1\}, \label{eq:deqE}
    \end{align}
\end{subequations}
where the sets $\mathbf{s}=\{s_i^j \mid i\in\{0,\dots,N-1\},\,j\in\{0,\dots,a\}\}$, $\mathbf{c}=\{c_i^j \mid i\in\{0,\dots,N-1\},\,j\in\{0,\dots,b\}\}$, $\mathbf{t}=\{t_i \mid i\in\{0,\dots,N-1\}\}$, $\tilde{x}(\cdot)$ and $\tilde{u}(\cdot)$ are piecewise continuous functions constructed using $s_i^j$ and $c_i^j$ as in~\eqref{eq:equation4A} and~\eqref{eq:equation4B} respectively. The evaluation mesh point $\rho_i^k$ denotes the $k^{\text{th}}$ quadrature point inside the $i^{\text{th}}$ interval, $\sigma_i^k$ being the quadrature weights for the $i^{\text{th}}$ interval and the constraint~\eqref{eq:deqE} ensuring that the smallest mesh is always above a certain threshold $t_{tol}$ normalized by the number of mesh intervals $N$. The internal supports $\tau_i^j$ and evaluation points $\rho_i^k$ depend on the mesh nodes $t_i$ according to a pre-defined relative distribution of nodes $\tau_{rel}^j$ (and $\rho_{rel}^k$ respectively) in each subdomain 
\begin{equation}
    \tau_i^j(t_i,t_{i+1})=\frac{t_{i+1}-t_i}{2}\tau_{rel}^j+\frac{t_{i+1}+t_i}{2}.
    \label{eq:discr}
\end{equation}
It is important to note that state continuity can be enforced at the mesh nodes $t_i$ by introducing the constraints of the form $s_i^a=s_{i+1}^0\ \forall i \in\{0,\dots,N-2\}$. However, in our formulation and implementation, we ensure that $\tilde{\chi}_i(t_{i+1})=\tilde{\chi}_{i+1}(t_{i+1}),\quad \forall i\in \{0,\dots,N-2 \},\label{eq:equationSCC}$ by using the same variable $s_{i}^a$ to represent both $\tilde{\chi}_i(t_{i+1})=s_{i}^a$ and $\tilde{\chi}_{i+1}(t_{i+1})=s_{i+1}^0$ and not include $s_{i}^0\ \forall i\in \{1,\dots,N-1 \}$ in the decision vector. 

\begin{figure}
    \centering
    \includegraphics[width=\columnwidth]{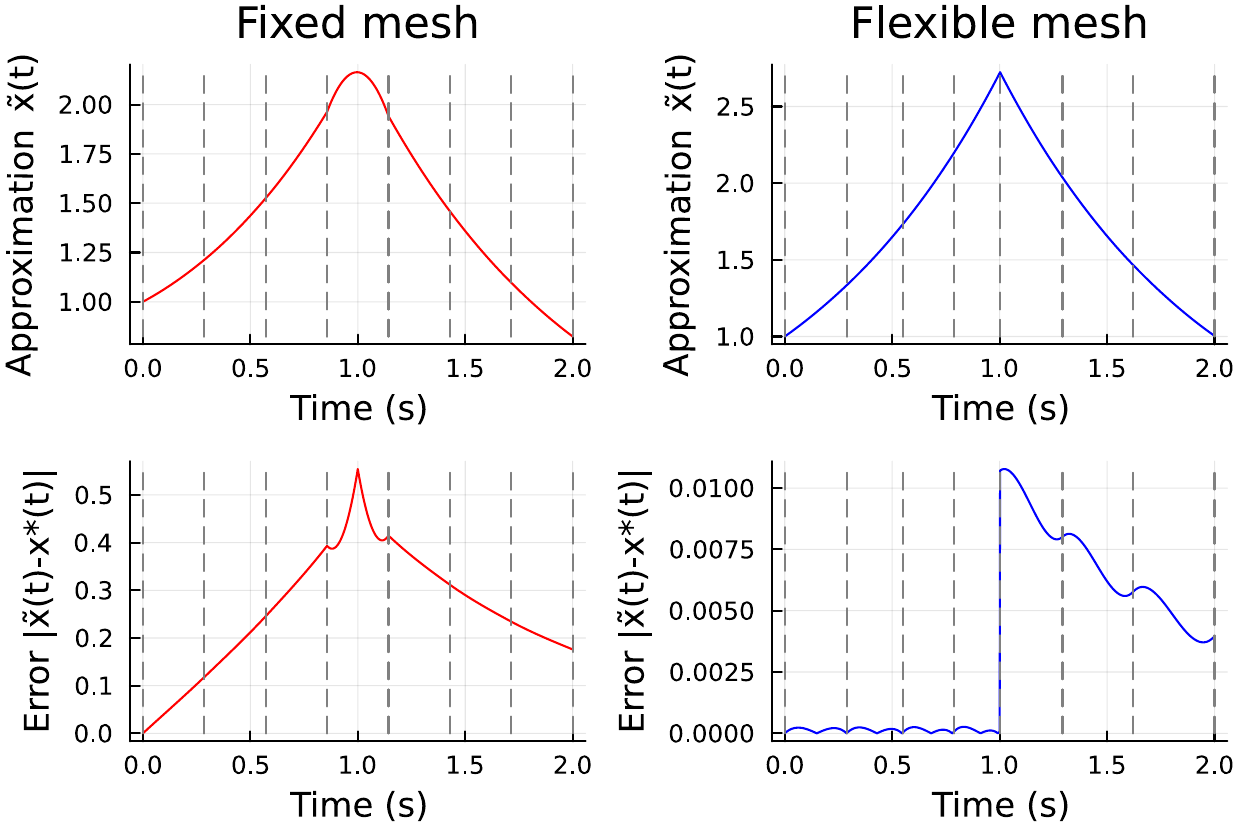}
    \caption{Numerical solution and absolute error plots of $\dot{x}(t)=-x(t)\cdot\text{sgn}(t-1)$ with initial condition $x(0)=1$ using $N=7$ intervals and polynomial degree $a=2$.}
    \label{fig:ODE_error}
\end{figure}

In order to solve differential equations, one may use $F$ to represent the differential equation system that needs to be solved. To showcase the performance of the flexible mesh for non-smooth differential equations, consider the following differential equation $F=\dot{x}(t)+x(t)\cdot\text{sgn}(t-1)=0,\, \forall t\in[0,2]$ with initial condition $x(0)=1$. The exact solution can be analytically computed (as $x^*(t)=e^t$ for $t<1$ and $x^*(t)=e^{2-t}$ if $t\geq 1$) and has a discontinuous derivative at $t=1$.  Figure~\ref{fig:ODE_error} shows the numerical solution $\tilde{x}(\cdot)$ obtained using piecewise quadratic approximation functions ($a=2$) and $N=7$ mesh intervals for fixed and flexible meshes. The solution obtained using the flexible mesh resembles the analytical solution much better, as shown in the subplots on the bottom row of Figure~\ref{fig:ODE_error} with the flexible mesh reducing the absolute error by more than an order of magnitude. In addition, it is also clear how a mesh node automatically located itself at $t=1$ in order to capture the point of non-smoothness.

\subsubsection{Flexible mesh for stiff problems}
\label{sec:3AC}

\begin{figure}
    \centering
    \includegraphics[width=0.8\columnwidth]{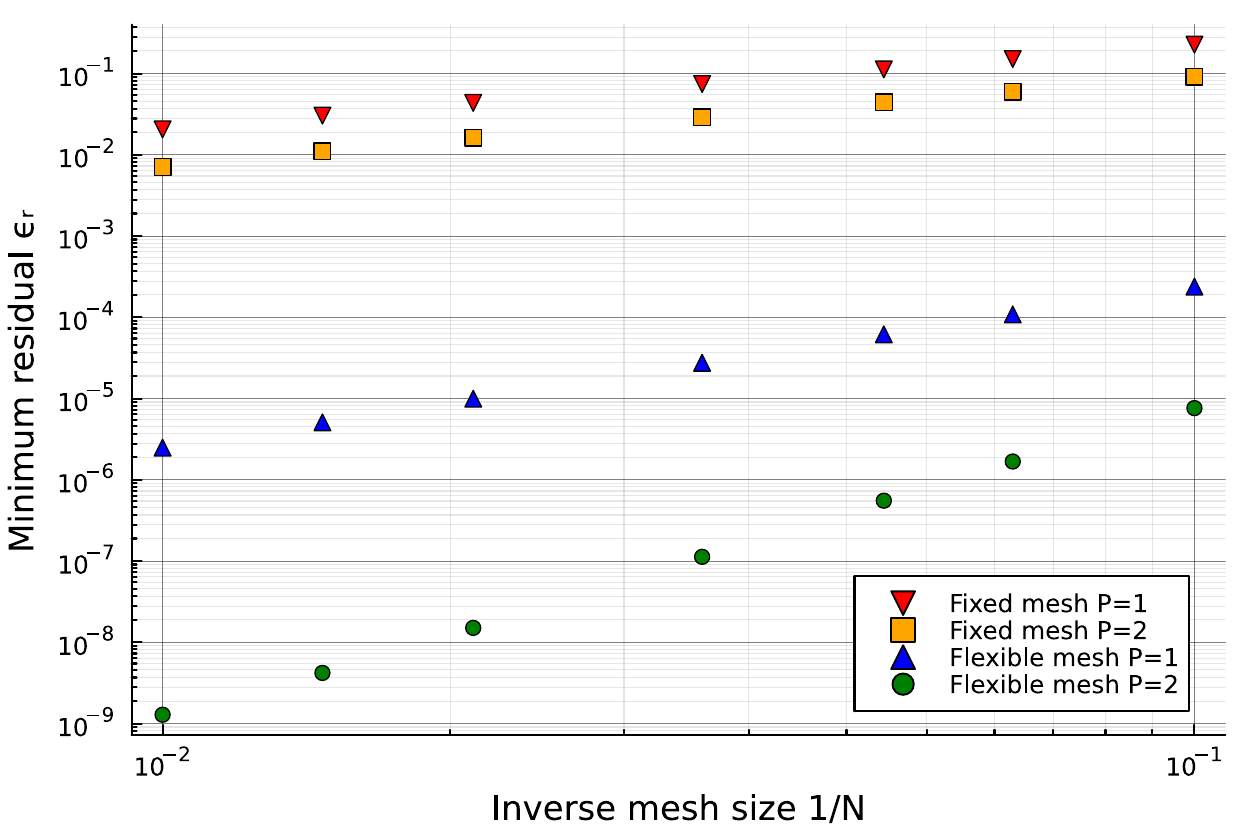}
    \caption{Convergence plot for a stiff Van der Pol system ($\nu=500$) where the integrated residual was defined as $\epsilon_{r}:=\int_{t_0}^{t_f}{\epsilon(t)}\diff t$ with the residual function $\epsilon(t):=\norm{ F(\dot{\tilde{x}}(t),\tilde{x}(t),\tilde{u}(t),t)}^2_2$}
    \label{fig:figure4}
\end{figure}

We have seen the benefits of using a flexible mesh if some of the trajectories are non-smooth, but one may be wondering whether it can still be useful if all of the trajectories are known to be smooth. To answer this question we will try to find a solution to the Van der Pol system $\dot{x}_1(t)=x_2(t),\ \dot{x}_2(t)=\nu[1-x_1^2(t)]x_2(t)-x_1(t)$ in the domain $t\in[0,2]$, with the initial conditions $x_1(0)=2$ and $x_2(0)=0$ and with $\nu$ being a constant parameter. This system is well-known for being stiff for high values of $\nu$. Most literature is concerned with order of convergence results and what the limiting behavior looks like in the vicinity of the solution (i.e.\ for a fine enough parametrization). In practice one should also look at how quickly the residual error decreases in early refinement iterations, since most algorithms start with a coarse mesh. To highlight why this matters, we will increase the stiffness of the system by making $\nu=500$ and compare the coarse mesh behavior (small $N$) for a fixed and a flexible mesh in Figure~\ref{fig:figure4}. As can be noticed, adding flexible nodes to a coarse mesh makes a bigger difference than adding fixed nodes. The flexible mesh starts capturing the dynamics using coarser meshes (as the slopes are much higher), thus achieving convergence for lower resolution meshes. Even if the order of convergence for a fixed mesh eventually tends to the same slope as the flexible mesh for high enough values of $N$ (since the solution to the Van der Pol system is continuous), the expected convergence order can only be observed for very fine meshes ($N>500$). In contrast, a flexible mesh can obtain an accurate solution using a smaller number of intervals.

\subsubsection{Solving optimal control problems}
\label{sec:3AD}

Once the original problem~\eqref{eq:equation2} has been discretized, a flexible mesh can be used to produce better numerical approximations $\tilde{x}(\cdot)$ and $\tilde{u}(\cdot)$ for state and input trajectories by solving the associated NLP

\begin{subequations}
    \label{eq:ocp_fm}  
    \begin{align}
        \min_{\mathbf{s},\mathbf{c},\mathbf{t}} \quad &  J(\tilde{x}(\cdot),\tilde{u}(\cdot),t_0,t_N) \label{eq:ocpfmA}\\
        \textrm{s.t. } &  \sum_{k=1}^{Q}{\sigma_i^k\cdot F_d(\dot{\tilde{x}}(\rho_i^{k}),\tilde{x}(\rho_i^{k}),\tilde{u}(\rho_i^{k}),\rho_i^k)^2} \leq \frac{\epsilon_{\text{max}}}{N}\label{eq:ocpfmB}\\
        &\eqref{eq:deqB},\,\eqref{eq:deqC},\,\eqref{eq:deqD},\,\eqref{eq:deqE} \label{eq:ocpfmC}
    \end{align}
\end{subequations}
where the constraint~\eqref{eq:ocpfmB} is enforced for each dimension $d\in\{1,\dots,N_F\}$ and each subdomain $i\in\{0,\dots,N-1\}$ and with evaluation nodes and weights $\rho_i^k$, $\sigma_i^k$. The Lagrange cost inside $J$ can be evaluated using quadrature rules, or integrated along with the dynamics by appending the state $x=[x^T, x_{N_x+1}]^T$ with dynamics of the newly added state defined as $\dot{x}_{N_x+1}=L(x(t),u(t),t)$ and initial condition $x_{N_x+1}(t_0)=0$. Note this increases the number of path constraints from $N_F$ to $N_{F+1}$.

For a low enough $\epsilon_{\text{max}}$ and a high enough $N$ the numerical solution $\tilde{x}(\cdot)$ and $\tilde{u}(\cdot)$ will converge to the exact solutions $x^*(\cdot)$ and $u^*(\cdot)$ with a total error defined as $\epsilon_t:=\int_{t_0}^{t_f}||\tilde{x}(t)-x^*(t)||_1+||\tilde{u}(t)-u^*(t)||_1\, dt$, where $||\cdot||_1$ represents the $L_1$ norm. A numerical example of using a flexible mesh for optimal control will be given in Section~\ref{sec:ch5}. In that section, a different error metric $\epsilon_t$ will be used to evaluate the accuracy of the solution a posteriori instead of the integrated residual $\epsilon_r$ to demonstrate the validity of our choice of error metric.

\subsection{Integrated residual transcription}
\label{sec:3B}

While flexible mesh design can be a useful tool in improving solution accuracy for non-smooth problems, it is important to employ an appropriate transcription method that accounts for errors in-between the nodal points. Carelessly employing a flexible mesh with collocation can result in nodes moving towards areas with small gradients where the solution might appear to be precisely captured because inter-nodal errors are neglected. As a consequence, the approximation error at the collocation points will be small, but the error in the regions ignored by the solver can be significant. To avoid such problems, the main transcription method proposed and employed in our work is an integrated residual formulation. The fundamental error metric used in this class of methods is the integrated residual $\epsilon_{r} \in \mathbb{R}$ defined as 
\begin{equation}
    \epsilon_{r}:=\int_{t_0}^{t_f}{\epsilon(t)}\diff t
    \label{eq:equation6}
\end{equation}
where the residual function considered is $\epsilon(t):=\norm{ F(\dot{\tilde{x}}(t),\tilde{x}(t),\tilde{u}(t),t)}^2_2$. In past work~\cite{nita2022fast,nita2022solving}, we have used $\epsilon_{r}$ directly to replace the constraints in~\eqref{eq:equation2B} by a single constraint $\epsilon_{r} \leq \epsilon_{\text{max}}$. While this is a valid alternative, in the current paper we instead enforce a similar constraint for each subdomain and for each dynamic equation in $F$ separately as
\begin{align}
\label{eq:equation7}
    \epsilon_i^{d}:=\int_{t_i}^{t_{i+1}}{F_d(\dot{\tilde{x}}(t),\tilde{x}(t),\tilde{u}(t),t)^2}\diff t\leq \frac{\epsilon_{\text{max}}}{N}\\
    \forall i \in \{0,\dots,N-1\},\ \forall d \in \{1,\dots,N_F\} \nonumber .
\end{align}
in order to avoid potential scaling issues and prevent loss of accuracy due to ill-conditioning. Both $\epsilon_r$ and $\epsilon_i^d$ are a consequence of the discretization. They can be interpreted as errors in representing constraints~\eqref{eq:equation2B} and are caused by inaccurate approximation of the state and input functions ($x$, $u$) by $\tilde{x}$ and $\tilde{u}$ (which are parameterized using a finite number of discretization nodes).

Since integration cannot be performed exactly, the common approach is to resort to quadrature rules to calculate the integrals numerically. This class of methods does not need to evaluate the approximation functions at the internal mesh nodes, but is rather a generalization that allows the construction of a different evaluation mesh (or quadrature mesh) $\rho_i^k$. Thus, it is easy to decouple the state mesh $\tau_i^j$ from the control input mesh $\mu_i^j$ since both $\tilde{x}$ and $\tilde{u}$ will be evaluated using the quadrature mesh $\rho_i^k$. An additional error $\epsilon_Q^{i,d}:=\Big|\epsilon_i^d-\sum_{k=1}^{Q}{\sigma_i^k \cdot F_d(\dot{\tilde{x}}(\rho_i^{k}),\tilde{x}(\rho_i^{k}),\tilde{u}(\rho_i^{k}),\rho_i^{k})^2\Big|}$ will appear as a consequence of approximating the integrals using numerical quadrature, with $\rho_i^{k}$ and $\sigma_i^k$, $k\in\{1,\dots,Q\}$ being the $Q$ quadrature nodes and weights for the interval $[t_i,t_{i+1}]$. Note the same quadrature mesh $\rho_i^k$ can be used to approximate the Lagrange cost term and a similar error metric $\epsilon_Q^{L}$ can be defined. However, the Lagrange cost can also be implemented by augmenting the state as previously explained. This approach would typically decrease the solve time and avoid the need of checking the quadrature error for both the dynamic constraints and the path cost. However, when augmenting the state, appropriate scaling might be required and the structure of the original problem~\eqref{eq:equation2} might be lost.

One of the main innovations brought about by integrated residual transcription is the possibility of decoupling the evaluation mesh ($\rho_i^k$) from the interpolation meshes ($\tau_i^j$ and $\mu_i^j$). This is relevant because it allows the user to have a very fine evaluation mesh in order to accurately represent and interpret the solution while maintaining a coarse interpolation mesh in order to keep the problem smaller in size, with a reduced number of decision variables, and thus improve the computational time. If the accuracy of the integration scheme is not sufficient, the quadrature errors $\epsilon_Q^{i,d}$ may be large, and consequently the numerically computed solution will not be valid. As a result, integrals will be reevaluated on a finer quadrature mesh (new $\rho_i^k$ for a higher $Q$) after a solution has been found. If the relative difference between the two is above a threshold $\varepsilon_{quad}$, the problem needs to be solved on a new, finer evaluation mesh. However, note that by refining the evaluation mesh, the decision vector does not increase in size since the number of decision variables $s_i^j$, $c_i^j$ only depends on the parametrization meshes $\tau_i^j$ and $\mu_i^j$ and not on the evaluation mesh $\rho_i^k$.     


\subsection{Feasibility driven optimal control}
\label{sec:3C}

In most practical control use cases, obtaining an accurate and feasible point faster is more desirable than achieving optimality. Consequently, using integrating residual transcription with a flexible mesh would enable one to obtain a feasible solution quickly and then iteratively improve the cost without resorting to many refinement iterations just to satisfy the dynamics constraints. Additionally, our algorithm allows the user to set a termination criterion based on the maximum computation time and is able to terminate early, returning a feasible point. This type of exit condition is ideal for most real-time model predictive control schemes where the requirement that the solve time be below the controller refresh rate is a crucial limitation.      
The method suggested in this paper is analogous to a simplex algorithm for solving linear programming problems, with phase one focusing on constraint satisfaction and phase two aiming to incrementally decrease the objective while remaining within the feasible set. In the same way that basic matrix operations can be used to navigate between different simplex vertices, in the non-linear case the current point can be used as an initial guess to warm-start the subsequent solves and significantly decrease the computation time.

\begin{figure}
    \centering
    \includegraphics[width=0.6\columnwidth]{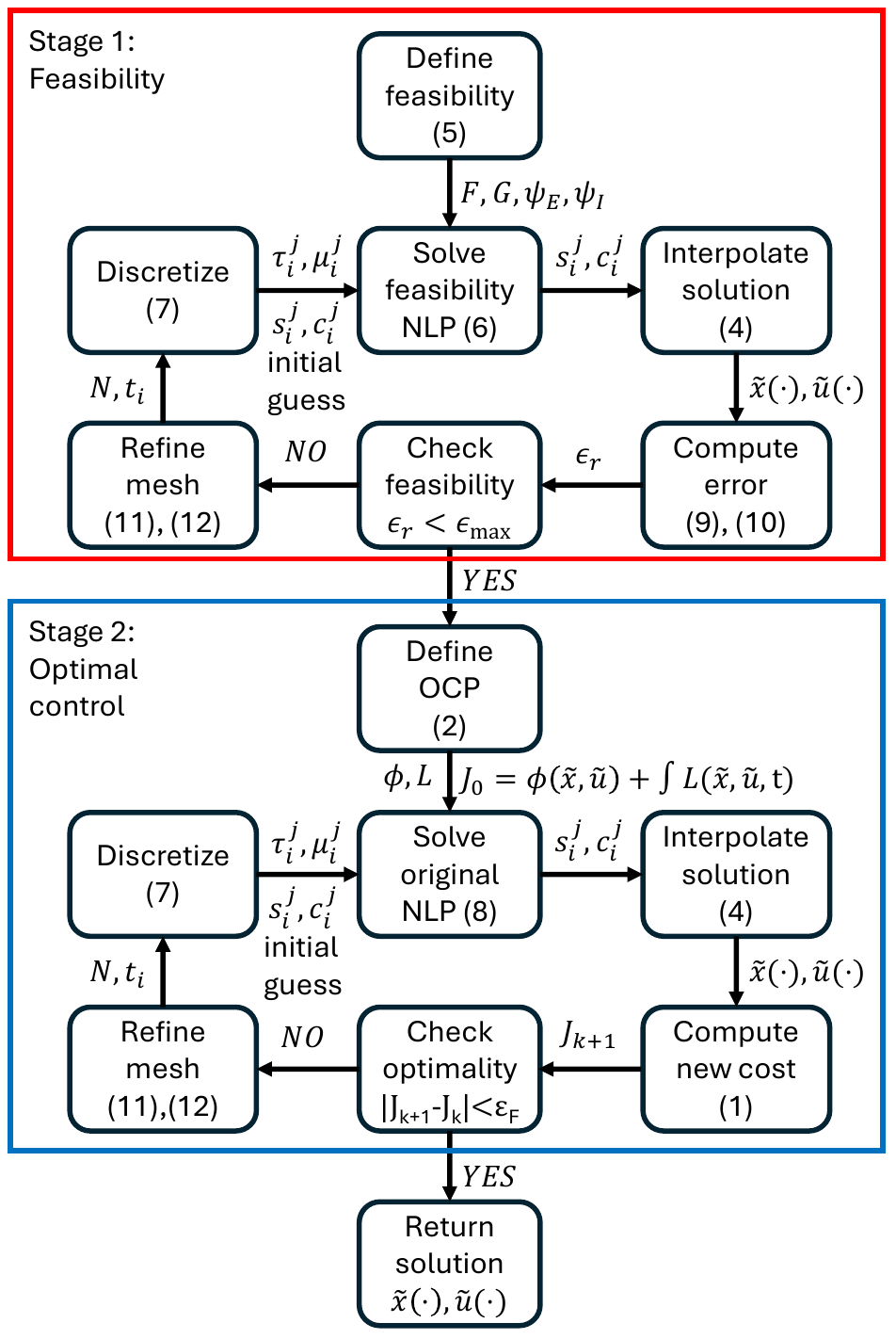}
    \caption{Feasibility driven solution approach: A schematic overview of the solution method. Note quadrature error was checked before feasibility and optimality in both stages, but it was omitted from the figure for enhanced readability.}
    \label{fig:figure10}
\end{figure}

A schematic of our solution method is presented in Figure~\ref{fig:figure10}. In order to solve the original problem given in~\eqref{eq:equation2}, we propose a two-step approach to ensure feasibility. Initially, a minimum residual problem is solved in order to obtain state and input trajectories that satisfy dynamics constraints. The discrete solution can then be interpolated using the chosen basis functions, and the dynamics error can be computed post-solve on a very fine mesh. In case the error is too high, the number of mesh intervals $N$ and/or the polynomial degrees $a$ and $b$ can be increased and a new refined mesh is generated.

Once the desired tolerance has been reached for every dynamic equation inside all the sub-intervals, we proceed to minimize the cost functional. Thus, we will solve~\eqref{eq:ocp_fm} using the solution of problem~\eqref{eq:diff_eq_flex_pb_def} as an initial guess. It is relevant to understand that while the solutions of problems~\eqref{eq:ocp_fm} and~\eqref{eq:diff_eq_flex_pb_def} may be significantly different and the computational time increases by adding the initial step, the benefit of this approach is that our algorithm can be terminated early with a feasible control sequence. Additionally, in practice, it is rather uncommon for the two problems to lead to vastly different state and input trajectories, and warm starting proves to be an efficient tool. In order to guarantee optimality, post-solve quadrature checks still need to be performed on both the dynamic constraint and the path cost, and the quadrature order increased if necessary.  

\subsection{Mesh refinement and algorithmic implementation}
\label{sec:3D}

In current state-of-the-art OCP solvers, an important step is mesh design and refinement~\cite{miller2021mesh}. In order to accurately capture the optimal control solution, mesh refinement is typically used to update the outer time mesh ($t_i$) as well as the inner sampling meshes ($\tau_i^j$ and $\mu_i^j$) until appropriate values for $N$, $a$ and $b$ have been found. Usually a full re-meshing is not necessary, and the coarse mesh solutions can be used as starting guesses in subsequent solves while only refining in the regions of interest. Although nodes will naturally migrate towards regions of high gradients, the guesses provided will impact the solver performance. When refining the mesh, we propose subdividing the largest intervals (highest $t_{i+1}-t_i$). This is because integrated residual transcription with a flexible mesh distributes the residual error more evenly across sub-domains, but larger intervals are more prone to quadrature error. Previous solutions can be used to estimate the convergence order and approximate the number of mesh nodes that need to be added to achieve the desired tolerance. This can be seen as numerically approximating the slopes in Figure~\ref{fig:figure2} and using them to estimate how many points are required for the specific desired accuracy. Using integrated residuals with a flexible mesh results in fewer nodes required to achieve a target accuracy, as well as fewer mesh refinement iterations when compared to both fixed- and flexible-mesh collocation. 

To generate the inner mesh, each interval $[t_i,t_{i+1}]$ is seeded with support points $\tau_i^j$ and $\mu_i^j$ according to~\eqref{eq:discr}. There are various well-established choices for the inner mesh design~\cite{trefethen1996finite} (choosing $\tau_{rel}^j$ and $\mu_{rel}^j$). In our implementation, we have chosen the outer mesh $t_i$ to be initialized as equispaced in the initial iteration and the internal grid was made using Chebyshev extreme points (also known as Chebyshev type 2). Parameterization and interpolation for states and inputs was performed using barycentric Lagrange interpolation polynomials as described in~\cite{amiraslani2016differentiation}. The mesh was refined according to 
\begin{equation}
    N=\frac{N_c^2}{N_{c-1}}^{\frac{\log_{10}{\frac{\epsilon_{\text{max}}}{\epsilon_{r,c}}}}{\log_{10}{\frac{\epsilon_{r,c}}{\epsilon_{r,c-1}}}}}
    \label{eq:newnodes}
\end{equation}
where $N$ is the new number of mesh intervals, $N_c$ and $N_{c-1}$ refer to the number of mesh intervals used in the current and previous refinement iteration respectively while $\epsilon_{r,c}$ and $\epsilon_{r,c-1}$ refer to the residual error obtained in the current and previous iteration. The first refinement iteration can be performed using any of the methods proposed in~\cite{betts2020practical}. The new nodes will be added proportionally in each interval 
\begin{equation}
    N_i=\floor{(N-N_c)\frac{t_i-t_{i-1}}{t_f-t_0}}
    \label{eq:nodes_dist}
\end{equation}
with $\floor{\cdot}$ being the floor function returning the largest integer smaller or equal to its argument. The remaining undistributed nodes will be assigned to the intervals one by one, starting with the largest. The distribution of new nodes inside existing subdomains is set to be equidistant. 

In the initial phase, the integrated residual minimization problem~\eqref{eq:diff_eq_flex_pb_def} is repeatedly solved until a feasible solution is obtained. After each solve, quadrature checks are performed to verify whether the solution is accurately captured by the evaluation mesh used, and if not, the number of quadrature points is increased. Numerical integration was performed using adaptive Gaussian quadrature as detailed in~\cite{quadgk}. If the quadrature is sufficient, the mesh nodes are added until the target residual tolerance is reached. Once problem~\eqref{eq:diff_eq_flex_pb_def} has been solved with the desired accuracy, the second phase solves problem~\eqref{eq:ocp_fm}. In our implementation, the path inequality constraints~\eqref{eq:deqB} are enforced at the support points $\tau_i^j$ in order to reduce the computational time and problem complexity. If inequality constraints are violated, it is easier to refine the mesh locally, rather than globally, without significantly increasing the computational burden. Alternatively one can exploit the properties of the chosen basis functions to ensure the inequality constraints are satisfied between the collocation points as well~\cite{vila2024tightly}, or define a different residual function that would account for path inequality constraints. The solutions obtained after each solve can be interpolated and used as starting points for the following iterations.  

\section{Numerical example}\label{sec:ch4}

\subsection{Van der Pol oscillator: singular control formulation}\label{sec:ch4A}

\begin{figure}
    \centering
    \includegraphics[width=0.8\columnwidth]{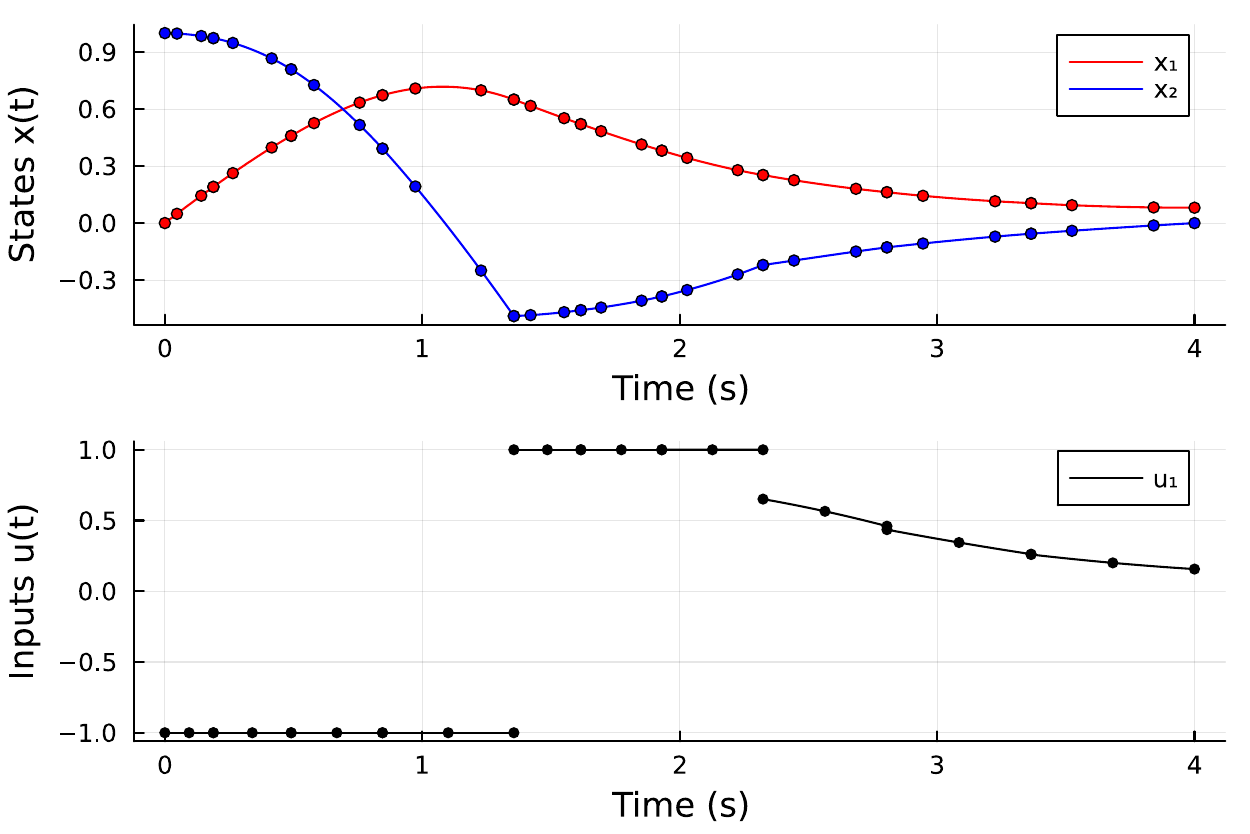}
    \caption{Optimal solution for Van der Pol control problem obtained using $N = 10$ flexible intervals, polynomial degrees $a = 3$ , $b = 2$, desired accuracy $\epsilon_{\text{max}}=10^{-6}$ and $t_{tol}=0.1$. The dots indicate inner node locations.}
    \label{fig:figure12}
\end{figure}

To demonstrate the effectiveness of our method,
we show the ability of our method to capture discontinuities and highlight its improved accuracy on the optimal control problem presented in~\cite{maurer_2007} with dynamics $\dot{x}_1 = x_2, \ \dot{x}_2 = -x_1 + x_2\left(p - x_1^2\right) + u$ based on the Van der Pol system, boundary conditions $x_1(0) = 0, \ x_2(0) = 1, \ t_0=0, \ t_f = 4$, input bounds $-1 \leq u(t) \leq 1$ and objective $J(x, u) = \frac{1}{2} \int_{0}^{t_f} \left(x_1^2 + x_2^2\right) dt$. The problem is known to have a bang-bang-singular structure, with two discontinuities located at non-trivial times, $t_{s1}=1.37$ and $t_{s2}=2.46$. These switching times are difficult to accurately capture by a numerical solver. To highlight how the flexible mesh is capable of capturing these discontinuities even when using a coarse low-resolution mesh, we have plotted the numerically obtained solution in Figure~\ref{fig:figure12}. This figure shows the state components $x_1$, $x_2$ and the control input $u$ obtained by implementing the algorithmic procedure described in Section~\ref{sec:3D}.

As can be seen, the control input is alternating between $-1$ and $1$, then follows a singular arc condition without having to enforce it explicitly. It is important to observe how interval boundaries $t_i$ automatically adjust in order to capture discontinuities in the control solution $u(t)$. Additionally, inside the subdomains where the exact solution is continuous the approximate solution curve looks smooth even for a low-accuracy mesh. Apart from the ability to capture discontinuities, in Figure~\ref{fig:figure12} one should also notice how the mesh automatically adapts and becomes denser close to the start of the simulation, since the gradient there is more rapidly changing than in other regions of the solution. The ability to capture regions of rapid change better than a fixed mesh is another key feature of our proposed flexible mesh.

\begin{figure} 
\centering 
    \begin{subfigure}[t]{0.45\columnwidth}  
    \centering 
    \includegraphics[width=\columnwidth]{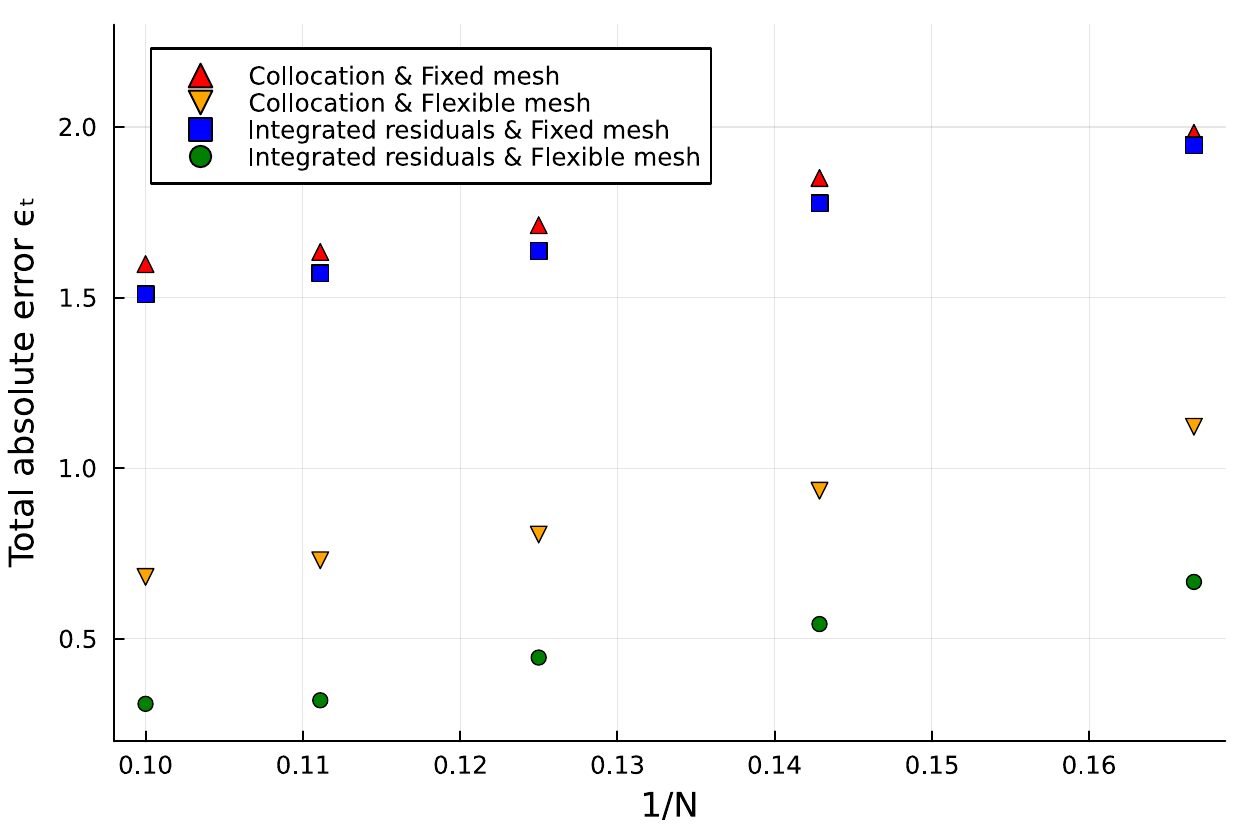} 
    \caption{} 
    \label{fig:figure13} 
    \end{subfigure}
     \hspace{0.01\columnwidth}
    \begin{subfigure}[t]{0.45\columnwidth} 
        \centering 
        \includegraphics[width=\columnwidth]{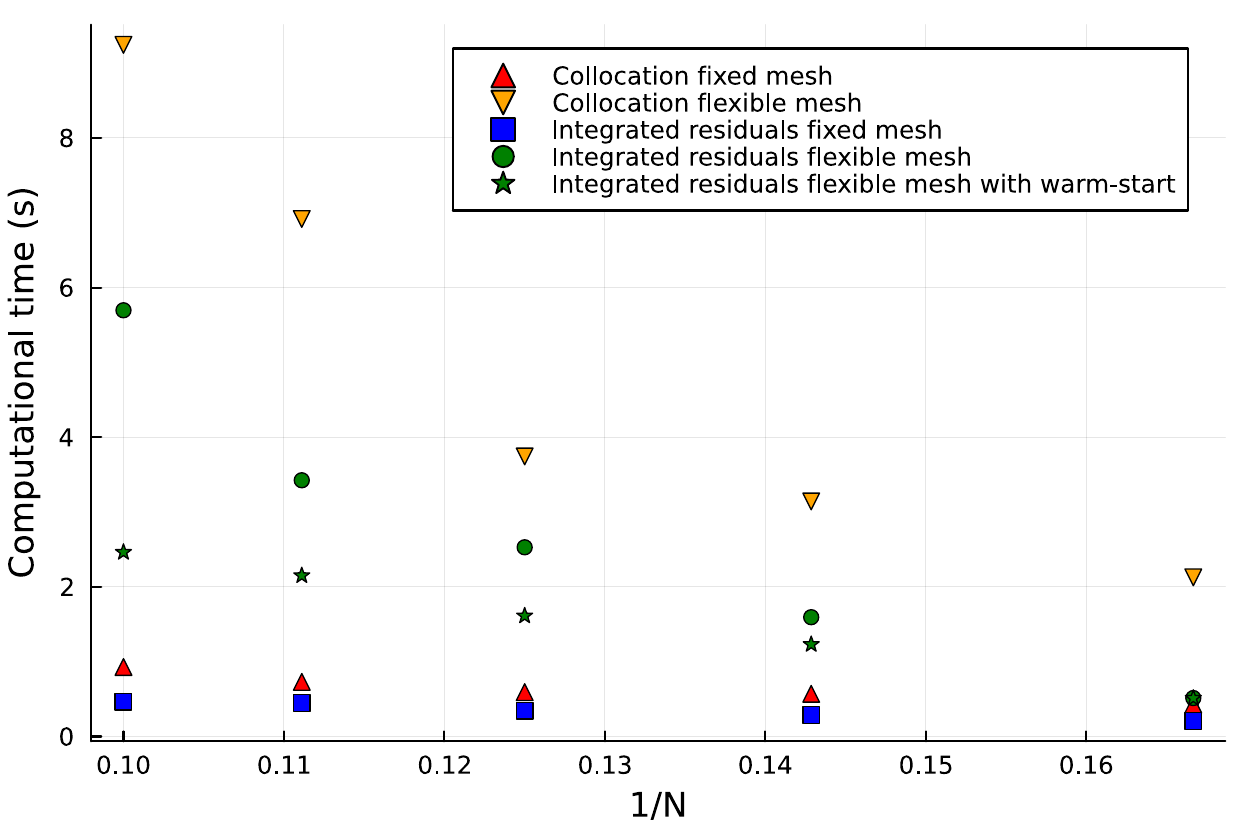} 
        \caption{} 
        \label{fig:figure14} 
    \end{subfigure}
    \caption{(a) Solution accuracy for Van der Pol problem as a function of average mesh size $1/N$. The polynomial degrees used were $a=2$ and $b=1$. (b) Computational time required to obtain a numerical solution to the Van der Pol control problem.}
\end{figure} 

By looking at Figure~\ref{fig:figure13} one can notice that the solution can be captured reasonably well even for a small number of mesh intervals if a flexible mesh is used. Using fewer intervals results into less memory being used, thus making the algorithm able to compute the solution on a smaller processor. In fact, the solution obtained using traditional collocation and fixed mesh can be almost three times less accurate than the solution computed using integrated residuals and a flexible mesh. Additionally, it is also obvious that the integrated residual method was proven to be more accurate than collocation for all the scenarios simulated, even if a fixed mesh was used (blue squares always being below the red triangles). By analyzing what happens to the solution as the mesh is refined, one would notice that increased accuracy would lead to adjusting the switching positions while maintaining the bang-bang-singular structure, getting closer to the exact solution. 

\begin{figure}
    \centering
    \includegraphics[width=0.8\columnwidth]{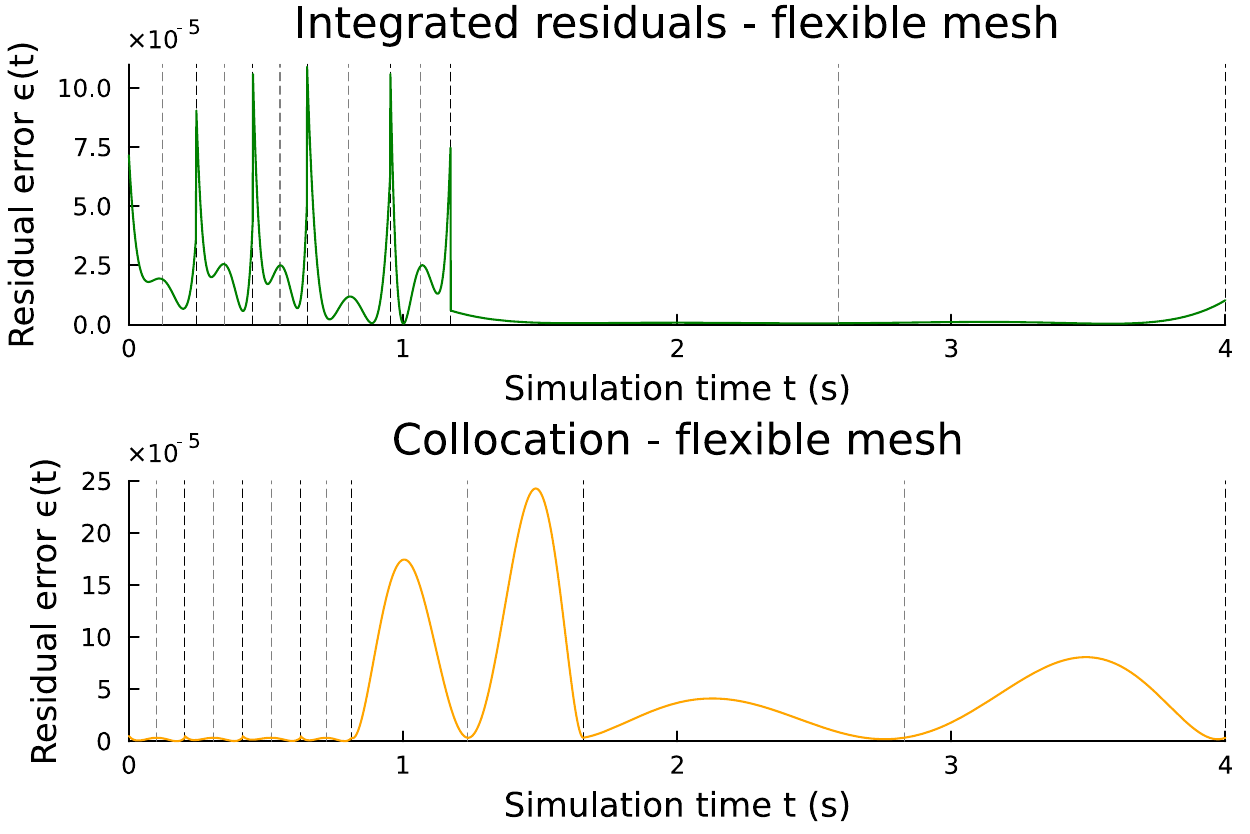}
    \caption{Residual error for flexible mesh integrated residual and collocation with $N=6$ mesh points and quadratic approximation $a=b=2$ for both state and input trajectories. Desired accuracy is set to $\epsilon_{\text{max}}=10^{-6}$ and $t_{tol}=0.1$}
    \label{fig:figure20}
\end{figure}

Figure~\ref{fig:figure20} illustrates why the integrated residual method is more appropriate for flexible meshing. The exact residual error (without numerical quadrature) is computed post-solve using the interpolated functions $\tilde{x}$ and $\tilde{u}$ and plotted for the two transcription methods. First, it is important to observe that if collocation transcription is used, the residual error is small at nodal points, but can be rather high in-between mesh points. In contrast, integrated residual transcription does not suffer from the same issue, and error peaks at the nodal points. Additionally, one should also observe that the maximum residual error for collocation is around $2.5$ times higher than that for integrated residual and is reached close to the location of the first discontinuity, meaning that the flexible mesh accurately captures the discontinuity only if used alongside integrated residual.

While being more accurate, the tradeoff one needs to consider when using a flexible mesh is the computational time required to obtain a solution. Firstly, by including mesh points as decision variables increases the dimension of the problem. Thus, it is expected that the NLP solver might take longer to find a solution than in the fixed-mesh case, since the decision space is higher-dimensional. Secondly, by adding mesh nodes as part of the problem definition, the structure of the original problem~\eqref{eq:equation2} will be altered. Additionally, using a flexible mesh makes the problem more coupled and introduces nonlinearities (since all the states and inputs will depend on the mesh node locations). As a consequence, the solution might become more sensitive to the initial guess. Figure~\ref{fig:figure14} shows the times taken to compute an approximate solution to the Van der Pol problem for a different number of mesh intervals $N$. Each experiment required to produce a data point in Figure~\ref{fig:figure14} was performed ten times, and the computational times reported are the average NLP solve time in seconds computed over the ten different runs. As can be seen, the blue squares are below the other data points, which means that in this example, for the attempted test cases, using an integrated residual transcription with fixed mesh is faster and more accurate (as can be deduced from Figure~\ref{fig:figure13}) than collocation over a fixed mesh. On the other hand, since the orange downward facing triangles are above all the other data points, this means that obtaining a numerical solution using a flexible mesh collocation is more time-consuming than all alternative formulations. 

To improve computational performance when using a flexible mesh for higher values of $N$, one should first attempt to solve the problem on a coarse mesh (low $N$) and implement the mesh refinement and warm starting strategy explained in Section~\ref{sec:3D}. To highlight the benefits of warm-starting, we have included in Figure~\ref{fig:figure14} the computational time required to compute an approximate solution using a flexible mesh with integrated residual and warm-start. This figure highlights the importance of warm starting when using a flexible mesh and performing multiple refinement iterations. As for the other cases, the reported times are an average of ten different runs. It can be seen that the benefit of warm starting also increases as the number of mesh intervals grows. In the best scenario simulated for $N=10$, the computational time for integrated residuals with flexible mesh was more than halved using warm start. 

\begin{figure}
    \centering 
    \includegraphics[width=0.8\columnwidth]{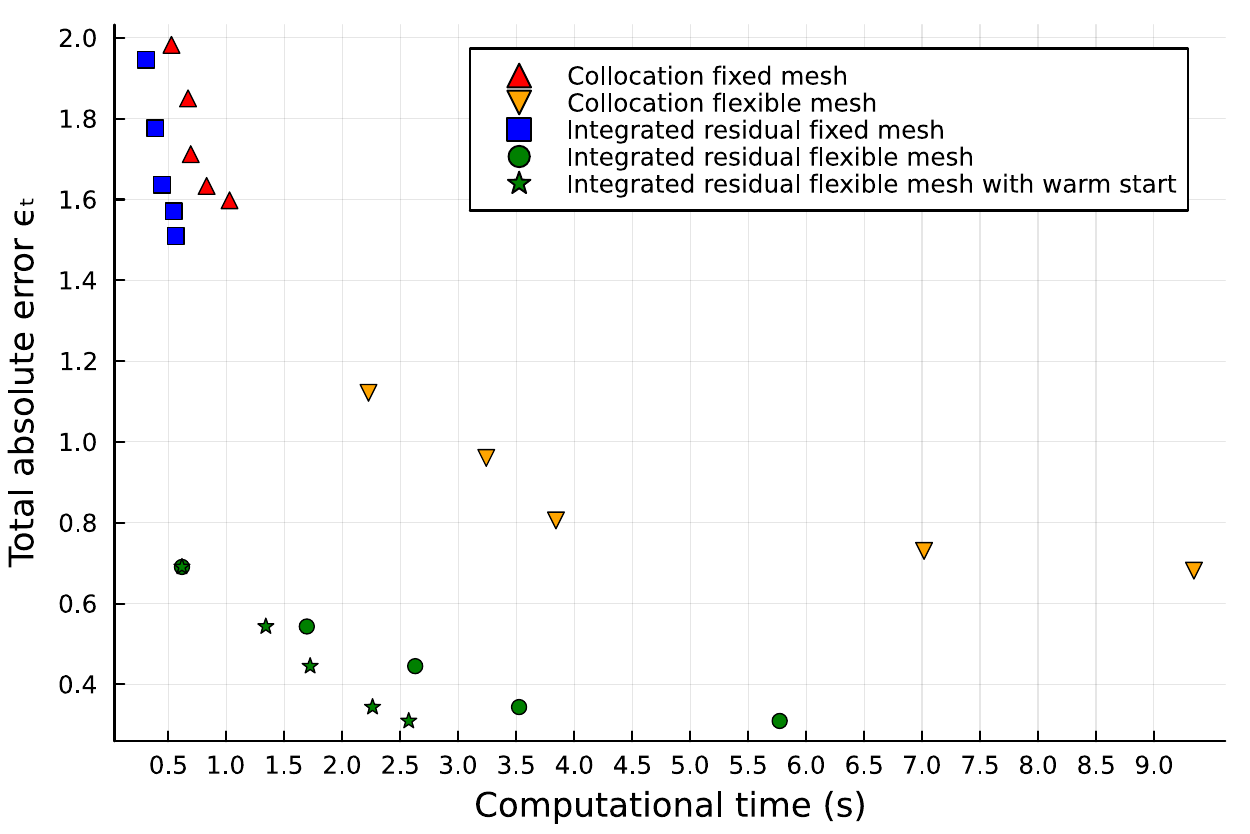} 
    \caption{Accuracy as a function of computational time obtained by combining Figures~\ref{fig:figure13} and~\ref{fig:figure14}. For the same computational time of 0.6\,s, using a flexible mesh and integrated residual leads to a solution that is twice as accurate as a fixed mesh collocation. } 
    \label{fig:figurecompt}         
\end{figure}

In Figure~\ref{fig:figurecompt} we have plotted the accuracy as a function of computational time for the Van der Pol problem. One can easily see that if integrated residual method with flexible mesh is used, the numerical solution that is more than two times more accurate than using the traditional fixed mesh collocation for the same computational time of about $0.6$ seconds.  

\subsection{Two-Link Robot Arm}\label{sec:ch4B}

\begin{figure}[tb]
    \centering
    \includegraphics[width=0.8\columnwidth]{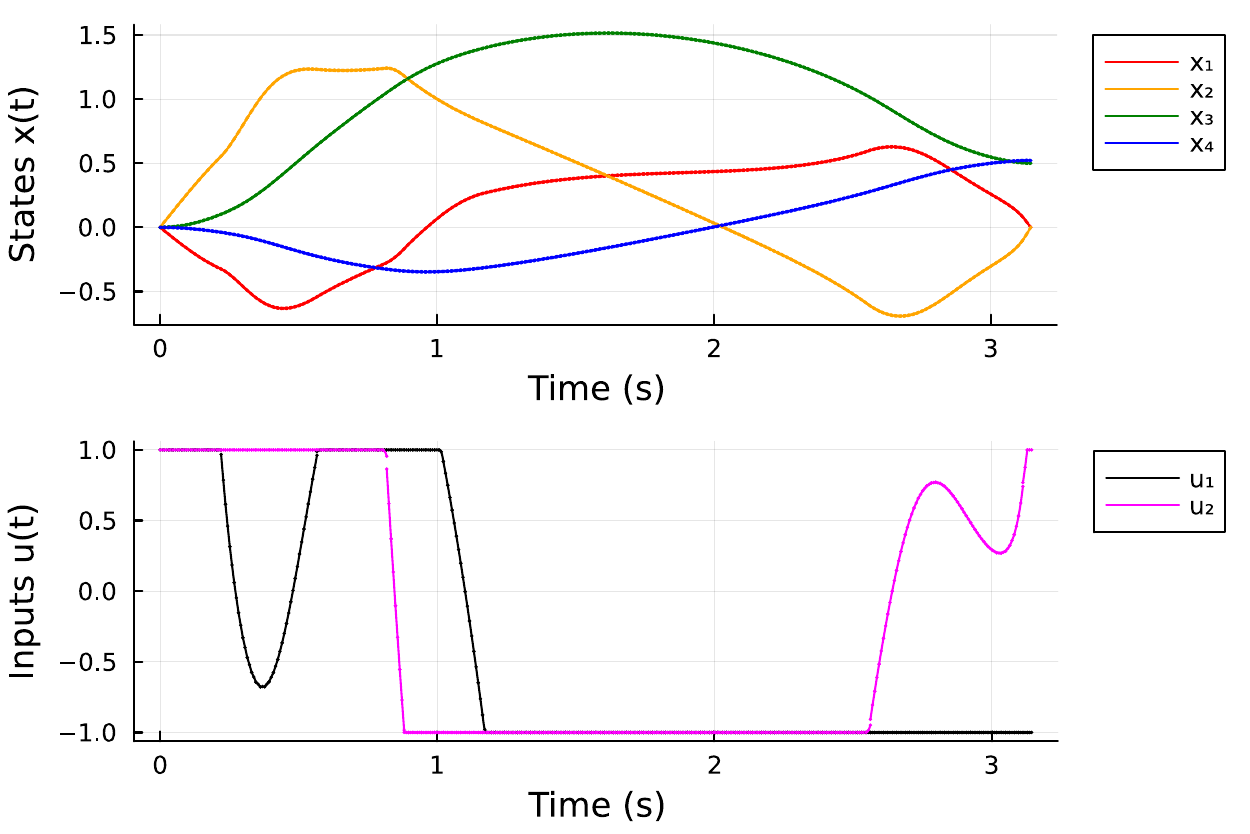}
    \caption{Numerically computed optimal control solution to the two link robot arm problem using integrated residuals and $N = 200$ fixed mesh intervals, polynomial degrees set to $a = 3$ , $b = 2$and desired accuracy $\epsilon_{\text{max}}=10^{-6}$. The small dots indicate the locations of the inner mesh nodes.}
    \label{fig:figure15}
\end{figure}

In this example, we will consider a two-link robot arm formulation as described in~\cite{nie2018iclocs2, luus2019iterative}. The computed solution is shown in Figure~\ref{fig:figure15}. It is important to note that even if input continuity is not enforced explicitly, the obtained solution will naturally converge towards the analytical solution, and the discontinuities that may appear for coarse meshes in the regions of rapid change will disappear as the mesh is refined.  

\begin{figure} [tb]
    \centering 
    \begin{subfigure}[t]{0.45\columnwidth} 
        \centering 
        \includegraphics[width=\linewidth]{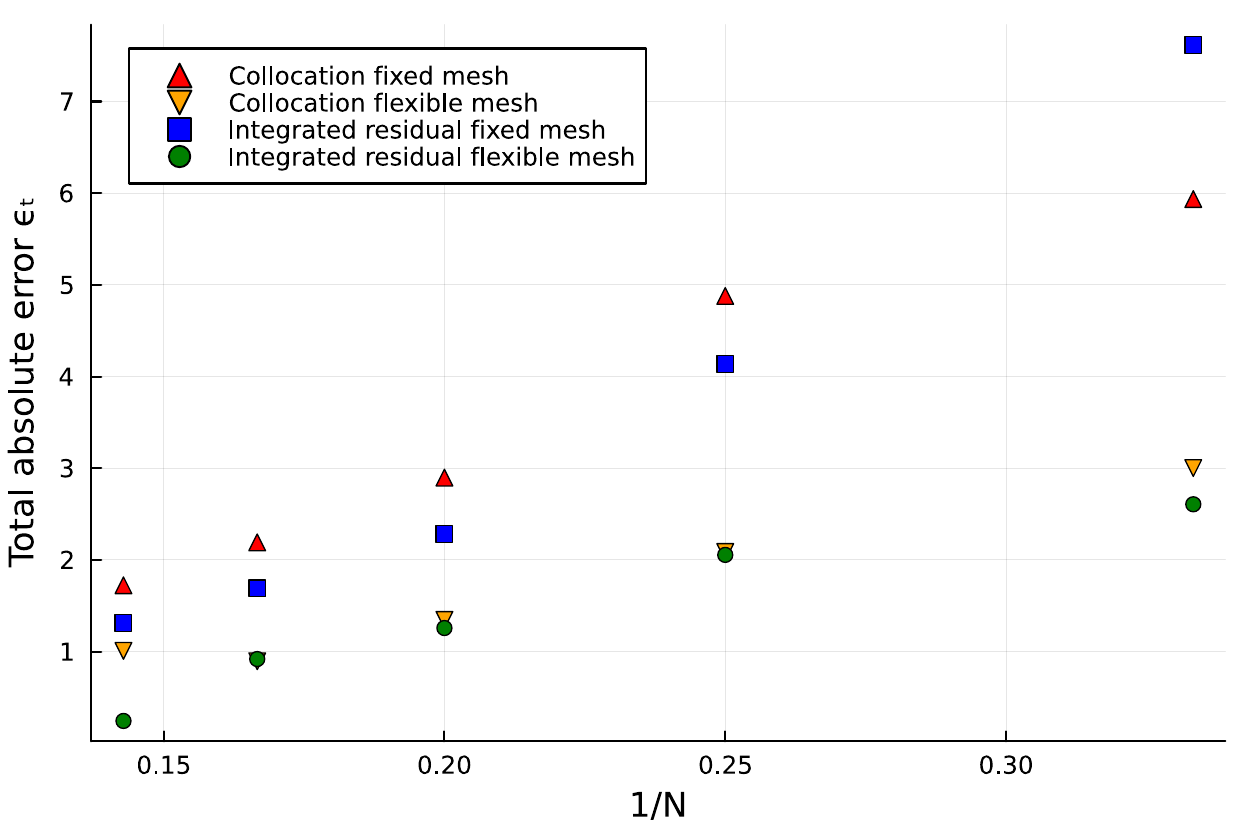} 
        \caption{} 
        \label{fig:figure19} 
    \end{subfigure}%
    \hspace{0.01\columnwidth}
    \begin{subfigure}[t]{0.45\columnwidth} 
        \centering 
        \includegraphics[width=\linewidth]{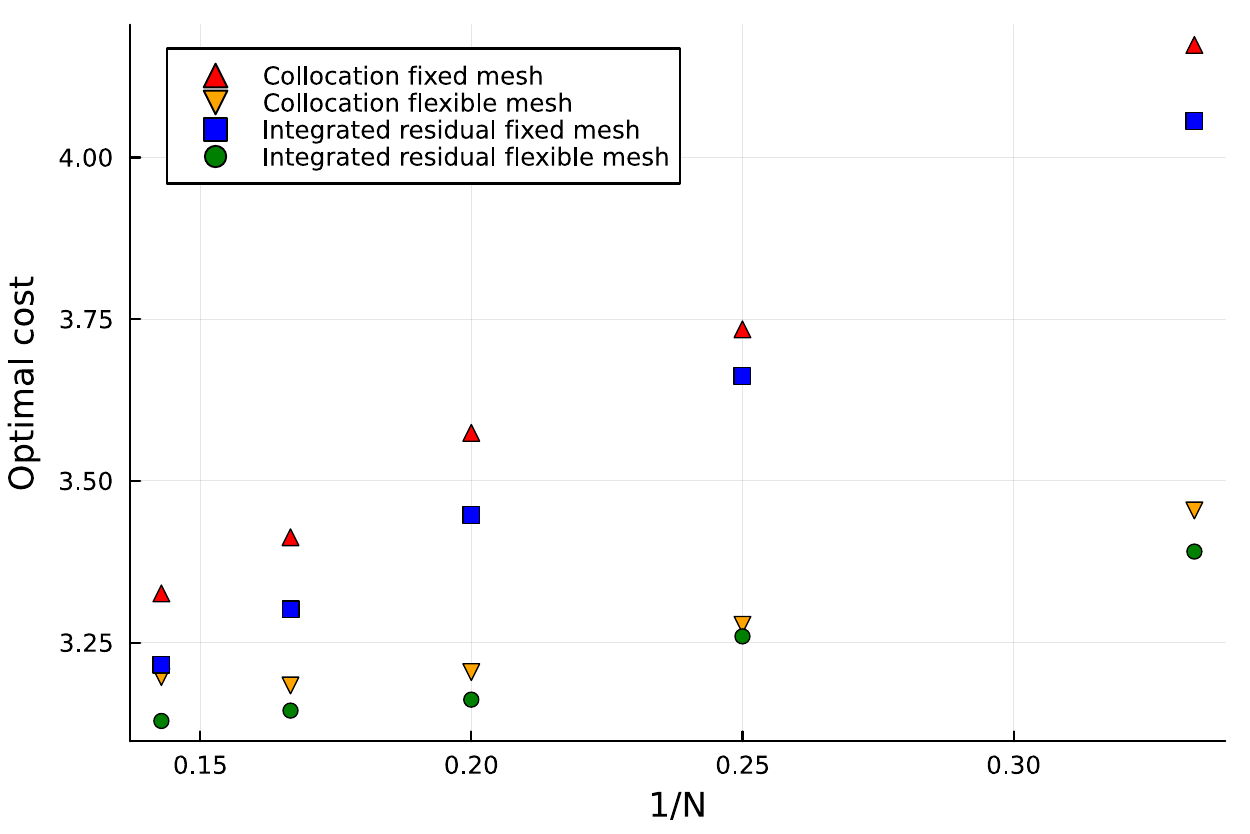} 
        \caption{} 
        \label{fig:figure18} 
    \end{subfigure} 
    \caption{(a) Solution accuracy for robot arm problem as a function of average mesh size $1/N$. The solution was computed for different numbers of mesh intervals ranging from $N=3$ to $N=7$. The polynomial degrees used to compute the solution were $a=3$ and $b=2$. The residual accuracy was $\epsilon_{\text{max}}=10^{-4}$. (b) Optimal cost for various discretization approaches as a function of mesh size for maximum residual tolerance $\epsilon_{\text{max}}=10^{-4}$ and polynomial degrees set to $a=3$ and $b=2$. } 
    \label{fig:figacc&ct} 
\end{figure}

Figure~\ref{fig:figure19} shows the total absolute error as a function of mesh size, for a fixed $\epsilon_{\text{max}}=10^{-4}$, which shows that a flexible mesh provides better accuracy for the same problem size even in the case of fully continuous solutions. The size of the problem is related to the memory allocation and the minimum size of the processor necessary to solve the problem. As a result, the integrated residual transcription joined with a flexible meshing idea is relevant for embedded processors, since smaller problems can be solved with the same accuracy. 

Figure~\ref{fig:figure18} illustrates the objective value against mesh size for the four discretization methods covered earlier with a residual accuracy goal of $\epsilon_{\text{max}}=10^{-4}$. Note that the numerical solutions obtained represent upper bounds for the optimal cost, which is achieved in the limit as the number of mesh points is increased. Thus, it becomes clear that the use of an integrated residual with a flexible mesh approach provides tighter upper bounds even for a small mesh size. This is consistent with the previous observation that the integrated residual improves the accuracy of the solution even for a small mesh size. It should also be noted that lowering the residual tolerance would result in a more accurate numerical representation of the dynamic constraint, and the solution obtained in the limit for a very fine mesh and a fixed $\epsilon_{\text{max}}$ is a lower bound to the exact analytical solution. From Figure~\ref{fig:figure18} it can be seen that as the mesh is refined for a fixed $\epsilon_{\text{max}}$, the objective value starts to plateau. The plateau shows that the maximum residual tolerance allowed $\epsilon_{\text{max}}$ needs to be lowered to further improve the approximation.

\begin{figure}[tb]
    \centering
    \includegraphics[width=0.8\columnwidth]{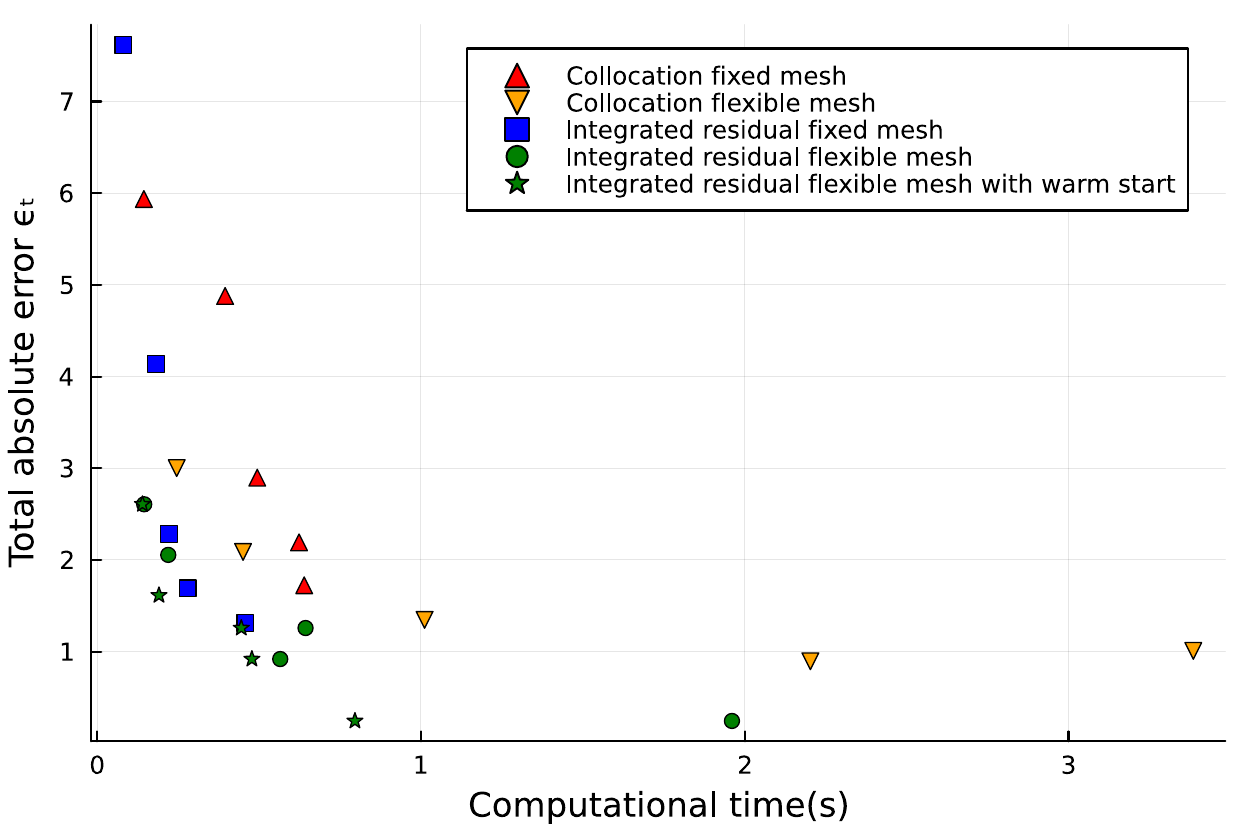}
    \caption{Solution accuracy for robot arm problem as a function of average mesh size $1/N$. The solution was computed for different numbers of mesh intervals ranging from $N=3$ to $N=7$. The polynomial degrees used to compute the solution were $a=3$ and $b=2$. The residual accuracy was $\epsilon_{\text{max}}=10^{-4}$.}
    \label{fig:RA_CT}
\end{figure}

Figure~\ref{fig:RA_CT} plots the accuracy as a function of computational time. Since this problem has a smooth solution, the computational time gains are not as significant as in the previous example, but the same trends can still be detected. That is, using a collocation discretization leads to an increase in computational time for the same solution accuracy. Additionally, using a flexible mesh can be useful and efficient for coarse meshes if fewer flexible nodes are used. As the size of the problem increases, the increase in computational cost eventually outweighs the improved accuracy brought about by flexible meshing. However, similarly to the previous example, the benefits of warm starting are still apparent. One can note that the accuracy will also eventually plateau as a result of using a fixed tolerance $\epsilon_{\text{max}}$. The plateau value for the flexible mesh collocation will be higher than the one for flexible mesh integrated residual, which highlights that a flexible meshing requires integrated residual transcription in order to achieve its full potential. 

\section{Conclusions}\label{sec:ch5}


Using a flexible mesh for numerically solving optimal control problems enhances the accuracy of the solution and identifies potential discontinuities. Furthermore, we have discussed that flexible meshing requires an integrated residual transcription to handle inter-nodal errors. Additionally, we pointed out that computing numerical approximations of the state and input trajectories involves a balance between accuracy and computational time. We have proposed a feasibility-driven method with warm starting to adapt the flexible mesh for real-time application. The numerical examples illustrated that an equal level of accuracy, compared to a fixed mesh, can be achieved using fewer flexible nodes and thus less memory. A flexible mesh is particularly beneficial for both problems with non-smooth solutions and stiff problems.

Future work can focus on providing theoretical guarantees on the order of convergence, as well as the impact of the mesh refinement strategy on computational time~\cite{neuenhofen2022quadratic}. Additional work can be done to develop a method for using a hybrid mesh with only a few flexible nodes. Currently, we have a systematic method to increase the quadrature order to ensure that the quadrature error remains minimal. However, future studies can focus on the numerical integration of discontinuous functions to reduce the number of iterations and better gauge the necessary quadrature order. Additionally, the selection of NLP solvers should be further investigated to identify more appropriate options for handling discontinuous functions.

\section*{Acknowledgments}

This research was supported by EPSRC under the Doctoral Training Grant EP/T51780X/1.

\bibliographystyle{plain}
\bibliography{your_bibliography_file}

\end{document}